\documentclass[11pt]{article}
\usepackage{amsfonts,amsmath}
\usepackage{epsfig}
\newcommand{\ba}{\begin{eqnarray}}
\newcommand{\ea}{\end{eqnarray}}

\newcommand{\Tr}{{\rm Tr}}

\newcommand{\Dslash}{\relax{\kern+.25em / \kern-.70em D}}

\newcommand{\Real}{\relax{\mathsf{\Gamma\kern-.35em R}}}
\newcommand{\Int}{\relax{\mathsf{Z\kern-.40em Z}}}

\newcommand{\gbar}{\kern1pt\overline{\kern-1pt g\kern-0pt}\kern1pt}
\newcommand{\mbar}{\kern2pt\overline{\kern-1pt m\kern-1pt}\kern1pt}
\newcommand{\obar}[1]{\kern3pt\overline{\kern-2pt #1\kern-0pt}\kern1pt}

\newcommand{\abar}{\kern1pt\overline{\kern-1pt a\kern-0.5pt}\kern1pt}

\newcommand{\D}{{\cal D}}
\newcommand{\la}[1]{\label{#1}}
\newcommand{\be}{\begin{equation}}
\newcommand{\ee}{\end{equation}}
\newcommand{\bi}{\begin{itemize}}
\newcommand{\ei}{\end{itemize}}
\newcommand{\rmi}[1]{{\mbox{\scriptsize #1}}}
\newcommand{\nr}[1]{(\ref{#1})}

\newcommand{\fr}[2]{{\frac{#1}{#2}}}

\renewcommand{\vec}[1]{{\bf #1}}

\newcommand{\Nv}{N_{v}}
\newcommand{\Ns}{N_{ s}}
\newcommand{\Pv}{P_{v}}
\newcommand{\Iv}{\vec{I_{v}}}
\newcommand{\Is}{\vec{I_{s}}}

\newcommand{\zz}%
 {{\mathbb{Z}}^4}
\renewcommand{\a}{a}    %{\alpha}
\renewcommand{\b}{b}    %{\beta}
\renewcommand{\c}{c}    %{\gamma}
\renewcommand{\d}{d}    %{\delta}
\newcommand{\RR}{{\rm I\kern -.2em  R}}
\newcommand{\eq}{Eq.~}

\def\lsi{\raise0.3ex\hbox{$<$\kern-0.75em\raise-1.1ex\hbox{$\sim$}}}
\def\gsi{\raise0.3ex\hbox{$>$\kern-0.75em\raise-1.1ex\hbox{$\sim$}}}

\newcommand{\gsim}{\mathop{\gsi}}

\makeatletter \@addtoreset{equation}{section} \makeatother
\renewcommand{\theequation}{\arabic{section}.\arabic{equation}}
\makeatletter
\renewcommand\section{\@startsection {section}{1}{\z@}%
                                   {-5.5ex \@plus -1ex \@minus -.2ex}% bfr-skip
                                   {2.3ex \@plus.2ex}%
                                   {\normalfont\large\bfseries}}
\renewcommand\subsection{\@startsection{subsection}{2}{\z@}%
                                     {-3.25ex\@plus -1ex \@minus -.2ex}%
                                     {1.5ex \@plus .2ex}%
                                     {\normalfont\normalsize\bfseries}}
\renewcommand\thesection {\@arabic\c@section}
\renewcommand\thesubsection   {\thesection.\@arabic\c@subsection}
\renewcommand{\@seccntformat}[1]{%
\csname the#1\endcsname.\hspace{1.0em}}
\makeatother
\begin{document}
\begin{titlepage}
\begin{flushright}
FTUV-07-0727 \\
IFIC/07-34\\
\end{flushright}
\begin{centering}
\vfill

\mbox{\Large\bf Finite-size scaling of the left-current correlator }
\vspace*{0.1cm}
\mbox{\Large\bf 
with non-degenerate quark masses}

\vspace*{0.8cm}

F.~Bernardoni\footnote{fabio.bernardoni@ific.uv.es}
and 
P.~Hern\'andez\footnote{pilar.hernandez@ific.uv.es}

\vspace*{0.8cm}

Dpto.\ F\'{\i}sica Te\'orica and IFIC, Edificio Institutos Investigaci\'on, \\
Apt.\ 22085, E-46071 Valencia, Spain

\vspace*{0.8cm}

{\bf Abstract}
 
\end{centering}
 
\vspace*{0.4cm}

\noindent
We study the volume dependence of the left-current correlator with non-degenerate quark masses to next-to-leading order in the chiral expansion. We consider three possible regimes: 
all quark masses are in the $\epsilon$-regime, all are in the $p$-regime and a mixed-regime where
the lighest quark masses satisfy $m_v \Sigma V \leq 1$ while the heavier $m_s \Sigma V \gg 1$.
These results can be used to match lattice QCD and the Chiral Effective Theory in a large but finite box in which the Compton wavelength of the lightest pions is of the order of the box size.
 We consider both the full and partially-quenched results. 
%% \vfill

%\noindent
%PACS numbers: 

%11.15.Ha, %        Lattice gauge theory
%11.30.Hv, %        Flavour symmetries
%11.30.Rd, %        Chiral symmetries
%12.38.Gc, %        Lattice QCD calculations
%12.39.Fe, %        Chiral Lagrangians
%\\
%Keywords:

\vspace*{1cm}
 
\noindent
%July 2006  %% \today

\vfill
 
\end{titlepage}

%%%%%%%%%%%%%%%%%%%%%%%%%%%%%%%%%%%%%%%%%%%%%%%%

\section{Introduction}

The first principles determination of the low-energy couplings of the Chiral Effective Theory, that describes the meson interactions at low momenta,  is one of the milestones of lattice QCD. This matching can only be carried out reliably close to the chiral limit, and this is often a limitation for lattice simulations, because the computational cost of lattice simulations increases very significantly with decreasing quark masses and increasing volumes.
 
Even though there has been important algorithmic progress in recent years, it seems quite difficult to reach the range of the $u$ and $d$ quark masses, at least  within the $p$-regime, i.e. satisfying the condition $M_\pi L \gg 1$.

The $\epsilon$-regime \cite{GL,N} has been advocated \cite{eregime} as an alternative to perform the matching, that a priori could be more economic in the sense that the quark masses can be taken to zero without increasing the box-size proportionally, since in this regime  $M_\pi L \leq 1$. Finize-size effects are large in this situation, but they are calculable within the Chiral Effective Theory in terms of the infinite-volume low-energy constants \cite{h}. Even though the truly chiral regime requires that the volume is scaled to infinity eventually, not just the quark mass to zero, the scaling with the volume at zero quark mass is more universal in the sense that it involves less low-energy couplings, since most of the operators that appear at higher orders in Chiral Perturbation Theory (ChPT)  include explicit powers of the quark mass. 

In particular it can be shown that in the $\epsilon$-regime only the leading order couplings $F$ and $\Sigma$ appear in 
two-point functions\footnote{Up to contact terms.}  at next-to-leading order (NLO)  of ChPT.

A number of correlation functions have been computed to NLO in the $\epsilon$-regime of ChPT. The two-point functions of scalar, pseudoscalar, vector and axial-vector correlators were presented in \cite{h}. In \cite{ddhj,currents} the same correlators were computed in quenched ChPT and also in the full theory, but in a fixed topological sector \cite{LS}.  Three-point functions relevant for determining 
the weak low-energy couplings were first presented in \cite{weak} both for the full and quenched theories.  The $\epsilon$-regimes has also  been recently applied to the study of baryon properties \cite{baryons}. These results have been used in various simulations to extract low-energy couplings mostly in the quenched approximation \cite{eregime}, but more recently also in unquenched simulations \cite{qcd}. 

In many of these computations quark masses were taken degenerate. The purpose of this paper is to present the results for the left-current correlator for non-degenerate quark masses. The methods developed here can be readily applied to other correlation functions. 

More concretely, we will consider the situation with $\Ns$ heavier quarks with common mass $m_s$ and $\Nv$ light quarks with common mass $m_v$. At this point one could imagine having three different situations:
\begin{itemize}
\item $m_{v/s} \Sigma V \gg 1$ : all quarks are in the $p$-regime. 

\item $m_{v/s} \Sigma V \ll 1$:  all quarks are in the $\epsilon$-regime. 

\item $m_v \Sigma V \leq 1$, $m_s \Sigma V \gg 1$: some quarks are in the $\epsilon$ and some in the
$p$ regime. It appears natural  to identify $m_v$ with the light quarks $u,d$ and $m_s$ with the $s$ quark mass in realistic simulations.
\end{itemize}

The explicit $\Nv$ and $\Ns$ dependences will be shown in such a way that the partial quenching of the 
$v$ or $s$ quarks can be easily done via the replica method \cite{replica}. Considering the partial quenching of the lighter quarks is interesting in the context of mixed-actions \cite{mixed}, where 
the valence and sea quarks are treated in different regularizations, for example with {\it overlap} valence quarks and Wilson sea quarks.

The structure of the paper is as follows. In section \ref{sec:full}, we present the results for the current correlator in the full theory, in the $p$, $\epsilon$ and mixed regimes. In \ref{sec:pq} we present the results for the partially-quenched theory, where the $v$ quarks are quenched, also in the three regimes.
Although the $\epsilon$ and $p$ regime results could have been obtained from earlier literature, we include them for completeness. The mixed-regime on the other hand involved a new method to separate the perturbative and non-perturbative modes, that will be discussed in detail.  In section \ref{sec:conclu} we present our conclusions and outlook. 

The results in this paper rely heavily on previous similar computations  in \cite{weak}. We refer the reader to those papers for further details of some intermediate steps.

\section{Full Theory Results}
\label{sec:full}

We start by considering $SU(N_s+N_v)$ Yang-Mills theory  with $N_s$ flavours with masses $m_s$ and $N_v$ flavours with masses
$m_v$. The quark part of the Euclidean continuum Lagrangian reads 
\be
 {L}_E = \sum_{r = 1}^{N_v} \bar \psi_r (\gamma_\mu D_\mu + m_v)\psi_r + \sum_{r = N_v+1}^{N} \bar \psi_r (\gamma_\mu D_\mu + m_s)\psi_r
 \;, 
\ee
where $r$ is a flavour index;
the Dirac matrices $\gamma_\mu$ are assumed normalised 
such that $\gamma_\mu^\dagger = \gamma_\mu$, 
$\{\gamma_\mu,\gamma_\nu\} = 2 \delta_{\mu\nu}$; 
$D_\mu$ is the covariant derivative and $N\equiv \Nv+\Ns$. We will consider external sources $J$ that have non-zero elements only in the $SU(\Nv)$ flavour subgroup. 

At large distances, the physics of QCD can be reproduced by chiral perturbation theory.  The leading order chiral Lagrangian reads
\ba
 \mathcal{L}_\rmi{ChPT} \!\! & = & \!\! \frac{F^2}{4} \Tr
 \Bigl[ \partial_\mu U \partial_\mu U^{\dagger} \Bigr] 
 - { \Sigma \over 2} \Tr
 \! \Bigl[ e^{i\theta} M U + U^{\dagger} M e^{-i\theta}\Bigr] 
 \;,
 \la{XPT} \la{LE}
\ea
where $U \in $ SU($N$). The mass matrix is diagonal with eigenvalues $(m_v,...,m_v,m_s,...,m_s)$
and $\theta$ is the vacuum angle. Apart from $\theta$,   
this Lagrangian contains two parameters, the pseudoscalar decay constant $F_{}$ and the 
chiral condensate $\Sigma$.  At NLO in the momentum expansion, additional operators appear in the chiral Lagrangian, with the associated low-energy constants 
$L_1,L_2,...$~\cite{gl2}. For a general $N$, the number of independent couplings is 11+2 \cite{bce}, but for 
$N=2$ and $N=3$, not all of them are independent and smaller subsets of 7+3 and 10+2 couplings respectively are commonly used in this situation\cite{gl2}. These couplings do not depend on the quark masses, but do depend on $N$. 

One of the simplest correlation functions that can be used to measure $F$ and is also sensitive to 
the NLO couplings $L_4, L_5, L_6$ and $L_8$ is the left-current two-point function. The numerical advantages of such correlator have been discussed in \cite{methods}.

In QCD, the left-handed flavour current can be formally defined as
\be
 J^a_\mu \equiv \bar\psi T^a \gamma_\mu P_- \psi 
 \;, \la{Jamu}
\ee
where $T^a$ is a traceless generator of the subgroup $SU$($\Nv$), and all colour, flavour, and spinor indices are assumed contracted. $P_- = (1-\gamma_5)/2$ is the left projector. 
Note that $J^a_\mu$ defined this way
is formally purely imaginary.\footnote{%
We use 
 this ``unphysical'' convention since it removes a number of 
 unnecessary overall minus signs from the ChPT predictions. 
 } 

The two-point correlation function between 
the left-handed currents, averaged over the spatial volume, now reads: 
\ba
  \Tr[T^aT^b]{C}(x_0) & \equiv & 
 \int \! {\rm d}^3 x\, 
 \Bigl\langle {J}^a_0(x) {J}^b_0(0) \Bigr\rangle .
 \la{Cqcd} 
\ea

On the ChPT side, the operator corresponding to~\eq\nr{Jamu} becomes, 
at leading order in the momentum expansion, 
\be
 \mathcal{J}^a_\mu = 
 \frac{F^2}{2} \Tr \Bigl[ T^a U \partial_\mu U^\dagger \Bigr] 
 \;.
\ee
The two-point correlation function $\mathcal{C}(x_0)$ is defined
(apart from contact terms) by
\be
 \Tr [T^a T^b] \, \mathcal{C}(x_0) = 
 \int \! {\rm d}^3 x \,  \Bigl\langle
 \mathcal{J}^a_0(x) \mathcal{J}^b_0(0) \Bigr\rangle
 \; . \la{Cxpt}
\ee

\subsection{The $p$-regime}
\label{sec:p}

In the $p$-regime, we express the outcome as a power series in $M^2/F^2$, 
where $M^2 \equiv 2 m \Sigma/F^2$ is the pseudoscalar mass. 
The power-counting rules for the $p$-regime are 
\be
 M \sim p \sim L^{-1}
 \;,
\ee
where $p$ is assumed a small quantity, $p\ll 4 \pi F$. 
The temporal extent $T$ can in principle
be small or large, as long as $T\gsim 1/(4\pi F)$. 
It follows from these assignments that the Goldstone field 
$\xi$, defined through $U = \exp(2 i \xi/F)$,    
behaves effectively as a small quantity, and can be expanded in. Here we have also 
set $\theta = 0$, as is usually done in the $p$-regime. 

Inserting the Taylor-series 
of $U$ into \eq\nr{LE}, the propagator in the quark basis becomes 
\ba
\Bigl\langle \xi_{\c\a}(x) \, \xi_{\d\b}(y) \Bigr\rangle  = 
 \fr12 \Bigl[\delta_{\c\b} \delta_{\d\a} G(x-y;M_{ab}^2) - 
 \delta_{\c\a} \delta_{\d\b} E(x-y;M_{aa}^2, M^2_{cc}) \Bigr]\;, \la{gen_prop}
 \ea
 where 
 \be
 G(x;M_{ab}^2) \equiv 
 \frac{1}{V} 
 \sum_{n \in \zz} 
 \frac{e^{i p \cdot x}}{p^2+M^2_{ab}}
 \;, \quad
 p \equiv (p_0,\vec{p}) 
 \equiv 2\pi\Bigl( \frac{n_0}{T}, \frac{\vec{n}}{L} \Bigr)
 \;.
 \label{g}
\ee
$V \equiv T L^3$ is the volume and $M_{ab}^2 = {\Sigma (m_a+m_b) \over F^2}$ is the mass of a meson  
constructed out of an $a$ and $b$-flavour quark, which in practice can be either $s$ or $v$.  On the other hand the singlet contribution is 
\be
E(x;M_{aa}^2, M_{cc}^2) \equiv 
 \frac{1}{V}
 \sum_{n \in \zz} 
 \frac{e^{i p \cdot x}}{(p^2+M_{aa}^2)(p^2+M_{cc}^2){F}(p)},
\label{e}
 \ee
 with
 \be
 F(p) \equiv \left[ \frac{\Ns}{p^2+M_{ss}^2}+  \frac{\Nv}{p^2+M_{vv}^2} \right],  
  \ee
 if $\Ns+\Nv\neq 0$. 
 
 The result for the left-current correlation function in the $p$-regime, after spatial integration over the source positions and up to contact terms, is:
\ba
 \mathcal{C}^{p}(x_0) & = & \frac{F^2}{2} \biggl\{ \left( 
 1 +{\Delta}_F \right) M_{vv}^2 P_v(x_0)
 -  \frac{\Nv}{F^2} \frac{{\rm d} G(0;M_{vv}^2)} {{\rm d} T} - \frac{\Ns}{F^2} \frac{{\rm d} G(0;M_{vs}^2)}{{\rm d} T}   \biggr. \nonumber \\
 & + &  
\biggl. {\Delta}_M  M_{vv}^2 \frac{{\rm d}}{{\rm d} M_{vv}^2} \Bigl[ M_{vv}^2 P_v(x_0) \Bigr]
 \biggr\} 
 \;, \la{Ct_p}
\ea
where
\ba
\Delta_F &=&  - \frac{\Ns}{F^2} G(0;M_{vs}^2) - \frac{\Nv}{F^2} G(0;M_{vv}^2) + \frac{8}{F^2}  \Bigl[\Ns M_{ss}^2 L_4+ M_{vv}^2 \left(\Nv L_4 + L_5\right) \Bigr], \\
\Delta_M  &=&  \frac{E(0;M_{vv}^2,M_{vv}^2)}{F^2} - \frac{8}{F^2} 
\Bigl[\left(\Ns M_{ss}^2 + \Nv M_{vv}^2 \right) (L_4 -2 L_6) + M^2_{vv}  (L_5 - 2 L_8)\Bigr]. \nonumber\\
\ea
The temporal dependence  (for $|x_0| \le T$) is contained in the function
\ba
  P_v(x_0) & \equiv & \int \! {\rm d}^3 \vec{x} \, G(x;M_{vv}^2) = 
 \frac{1}{T} \sum_{p_0} \frac{e^{i p_0 x_0 }}{ p_0^2 + M_{vv}^2 }
 =  \frac{\cosh[M_{vv}(T/2 - |x_0|)]}{ {2 M_{vv}} \sinh[M_{vv}T/2]}
 \; . 
\ea
%\be
 %G_V(M^2) = 
 %\frac{1}{(4\pi)^2} 
 %\int_0^\infty \frac{{\rm d}\lambda}{\lambda^2} e^{-\lambda M^2}
 %% \sum_{n \neq 0}
 %\sum_{n \in \zz} 
 %\Bigl(1 - \delta^{(4)}_{n,0} \Bigr)
 %\exp \Bigl[
 %-\frac{1}{4\lambda} \Bigl( 
 %T^2 n_0^2 + L^2 
 % \sum_{i = 1}^3 n_i^2 
% |\mathbf{n}|^2
 %\Bigr) 
 %\Bigr]
 %\;.
 %\la{GV}
%\ee
%For $M V^{\fr14} \gg 1$, the finite-volume effects are exponentially small, and we can set $G_V = 0$.

Up to the second term, which is a constant finite-volume effect, the NLO propagator has the same temporal dependence as the LO, if the decay constant and pseudoscalar mass squared are scaled by a relative correction given by $\Delta_F$ and $\Delta_M$ respectively:
\ba
F_{NLO}^2 = F^2 ( 1 + \Delta_F ), \;\; M^2_{NLO} = M_{vv}^2 ( 1 + \Delta_M).
\ea
 These results agree with those obtained by Gasser and Leutwyler in infinite volume \cite{gl2} and finite volume \cite{GL} for $N_v=2$ and $N_s=1$. Finite volume corrections to $F$ and $\Sigma$ have been also obtained to two-loops \cite{cdh}. 

The finite volume corrections can be isolated by
\be
 G(0;M^2) \equiv G_\infty(M^2) + G_V(M^2) \; \quad  E(0;M^2,{M'}^2) \equiv E_\infty(M^2,{M'}^2) + E_V(M^2, {M'}^2) 
 \;,
\ee
where $G_\infty, E_\infty$ are the infinite-volume closed propagators, where instead of momentum sums in eqs.~(\ref{g}) and (\ref{e}) there are 
integrals \footnote{%
  The UV divergences of these quantities  for $d\approx 4$ cancel against
  those~\cite{gl2} in the $L_i$'s as expected.}.
%\be
% G_{\infty}(M^2) \equiv \int \! \frac{{\rm d}^d p}{(2\pi)^d} 
 %\frac{1}{p^2 + M^2}  \;  E_{\infty}(M^2_{vv},M_{vv}^2) \equiv \int \! \frac{{\rm d}^d p}{(2\pi)^d}  
 %\;,  
%\ee
The (finite) functions $G_V(M^2)$ and $E_V(M^2, {M'}^2)$  incorporate the volume
dependence~\cite{hal}\footnote{%
  In Ref.~\cite{hal}, the function $G_V$ was denoted by $g_1$. 
  }.

The limit in which $m_s \rightarrow m_v$, we of course recover the degenerate mass result of \cite{weak}.

\subsection{The $\epsilon$-regime}
\label{sec:e}

We consider now the case where all quark masses satisfy $m_{s/v} \Sigma V \leq 1$. The results for the correlator obtained in a $\theta$-vacuum in \cite{h} and in a fixed-topology in \cite{currents,weak} are valid  for non-degenerate quark masses:
\be
 \mathcal{C}^{\epsilon}(x_0) 
 = \frac{F^2}{2 T}
 \biggl[
 1 + \frac{N}{F^2}\biggl(
 \frac{\beta_1}{\sqrt{V}} - \frac{T^2 k_{00}}{V} \biggr)
 + \frac{2 T^2 }{F^2 V} \mu\sigma^{(N_s,N_v)}_\nu(M) h_1(\hat x_0 ) 
 \biggr] 
 \;, \la{Ct_eps}
\ee
where $\hat{x}_0 = x_0/T$.  The only non-trivial mass dependence is in the function $\mu\sigma^{(\Ns,\Nv)}_\nu(M)$:
\ba
\mu\sigma^{(N_s,N_v)}_\nu(M) \equiv \int_{U(N)} ~dU ~{\mu_v \over 2 \Nv}~{\rm Tr}\left[ \Pv U + U^\dagger  \Pv\right] ~(\det U)^\nu ~\exp\left({\Sigma V \over 2} {\rm Tr}\left[ M U + U^\dagger M \right] \right),
\label{musigma}
\ea
where $\Pv$ is the projector onto the sector of masses $m_v$, and $\mu_v \equiv  m_v \Sigma V$.

The constants $\beta_1$ and $k_{00}$ are related to the (dimensionally regularised) value of 
\be
 \bar G(x, M^2) \equiv \frac{1}{V} 
 \sum_{n \in \zz }
 \Bigl(1 - \delta^{(4)}_{n,0} \Bigr) \frac{e^{i p\cdot x}}{p^2+M^2} 
 \;, 
 \la{Gx}
\ee
by
\be
 \bar G(0,0) \equiv -\frac{\beta_1}{\sqrt{V}} \;, \quad
 T \frac{{\rm d}}{{\rm d} T} \bar G(0,0) \equiv \frac{T^2 k_{00}}{V} 
 \;. \la{beta1}
\ee
Introducing $\rho \equiv T/L$ and 
\ba
 \hat \alpha_p(l_0,l_i) & \equiv & 
 \int_0^1 \! {\rm d} t\, 
 t^{p-1} 
 \Bigl[
 S\Bigl( {l_0^2} / {t} \Bigr) 
 S^3\Bigl( {l_i^2} / {t} \Bigr) - 1 
 \Bigr] 
 \;, \la{alphap}
\ea
where $S(x)$ is an elliptic theta-function, 
$S(x) = \sum_{n=-\infty}^{\infty} \exp(-\pi x n^2)
= \vartheta_3(0,\exp(-\pi x))$, 
a numerical evaluation of these coefficients
is allowed by (see, e.g., Refs.~\cite{hal,h})
\ba
 \beta_1 & = & 
 \frac{1}{4\pi}
 \Bigl[ 2 - 
 \hat\alpha_{-1}\Bigl( 
 \rho^{\fr34},\rho^{-\fr14}
 \Bigr)
 -
 \hat\alpha_{-1}\Bigl( 
 \rho^{-\fr34},\rho^{\fr14}
 \Bigr)
 \Bigr]
 \;, \\
 k_{00}  & = &  
 \fr1{12} - \fr14\sum_{\vec{n}\neq \vec{0}}
 \frac{1}{\sinh^2(\pi \rho |\vec{n}|)}
 \;.
\ea
The function $h_1(\tau)$ appearing in~\eq\nr{Ct_eps} 
reads (for $|\tau|\le 1$)
\ba
 h_1(\tau) & \equiv & \frac{1}{2}  
 \left[\left(|\tau| - {1 \over 2}\right)^2 - {1 \over 12}\right]
 \;. \la{ph1} 
\ea
 
 The integral of eq.~(\ref{musigma}) for non-degenerate quark masses can be written in terms of a 
functional derivative:
\ba
\mu\sigma^{(N_s,N_v)}_\nu(M) = {m_v \over \Nv}~{1 \over Z^{(N_s,N_V)}_\nu(M_J)}~ \left.\frac{\partial Z^{(N_s,N_v)}_\nu(M_J)}{\partial J}\right|_{J=0}, 
\ea
where
\ba
 Z^{(N_s,N_v)}_\nu(M_J) \equiv  \int_{U(N)} ~(\det U)^\nu ~\exp\left({\Sigma V \over 2} {\rm Tr}\left[ M_J U + U^\dagger M_J \right] \right),
 \label{znu}
\ea 
where $M_J$ is the block-diagonal matrix:
\ba
M_J = \left(\begin{array}{cc}
             (m_v+J) \vec{I_v}  &0\\
             0  & m_s \vec{I_s} \\
\end{array}\right).
\ea
and $\Iv$ and $\Is$ are the identity matrices in the $v$ and $s$ subgroups respectively.

The functional $Z^{(\Ns,\Nv)}_\nu$ is known in terms of modified Bessel functions \cite{brower, LS}
\ba
Z^{(N_s,N_v)}_\nu(M_J) = C_\nu \frac{\det\left[  \mu_i^{j-1} I_{\nu+j-1}(\mu_i)\right]_{i,j=1,...,N}}{\prod_{j > i=1,..,N} (\mu_j^2 - \mu_i^2)},
\label{znufull}
\ea
where $I_n(x)$ is the modified Bessel function and $\mu^2_i$ are the eigenvalues of the matrix $M_J^\dagger M_J$ multiplied by $(\Sigma V)^2$.

In our case we just have to consider two distinct eigenvalues $m_s^2$ and $(m_v+J)^2$. As a useful example we consider the case with $N_v=2$ and $N_s=1$, corresponding to the $2+1$ flavour QCD:
\ba
Z^{(1,2)}_{\nu}(M) &=& {1\over 2}   I_\nu(\mu_s) \left[ I_\nu(\mu_v)^2 - I_{\nu+1}(\mu_v) I_{\nu-1}(\mu_v)\right] \nonumber\\
&+& \frac{I_\nu(\mu_v)}{\mu_s^2 -\mu_v^2} \left[ \mu_v I_{\nu+1}(\mu_v) I_\nu(\mu_s) - \mu_s I_\nu(\mu_v) I_{\nu+1}(\mu_s)\right] ,
\ea
while 
\ba
\mu\sigma^{(1,2)}_{\nu}(M) =  {\mu_v \over 2 Z^{(1,2)}_\nu(M)}    \frac{d}{d\mu_v} Z^{(2,1)}_\nu(M).
\ea
Another interesting case is that of $N_v=2$ and $N_s=0$, corresponding to $2$ flavour QCD. In this case we have 
\ba
Z^{(0,2)}_{\nu}(M) =   {1\over 2} \left[ I_\nu(\mu_v)^2 - I_{\nu+1}(\mu_v) I_{\nu-1}(\mu_v)\right] ,
\ea
while 
\ba
\mu\sigma^{(0,2)}_{\nu}(M) =   \frac{ I_{\nu+1}(\mu_v) I_{\nu-1}(\mu_v)}{\left[ I_\nu(\mu_v)^2 - I_{\nu+1}(\mu_v) I_{\nu-1}(\mu_v)\right] }.
\ea

\subsection{The mixed-regime}
\label{sec:m}

Now we turn to the most complicated case of the mixed regime. In this case, some quarks 
are in the $\epsilon$ and some in the $p$ regime and therefore a different factorization of zero and non-zero modes is needed. As in the previous sections we start by considering the full theory case 
with the $v$ and $s$ quarks are both unquenched. 

The power-counting for this regime is
\ba
m_v \sim \epsilon^4 \; \; m_s \sim p^2 \sim L^{-2} \sim \epsilon^2 .
\label{powerc}
\ea

The inspection of the $p$-regime propagator of eq.~(\ref{gen_prop})  shows that the modes that become massless in the $m_v \rightarrow 0$ limit are those corresponding to the generators of $SU(N_v)$. Therefore a factorization that would treat the zero-momentum modes of these fields non-perturbatively is 
\ba
U = \left(\begin{array}{cc}
U_0 & 0 \\
0   & \Is \end{array}\right) \; \exp\left( {2 i \xi \over F}\right),
\label{factor_mixed}
\ea
where $\vec{I_s}$ is the identity matrix in the $s$ sector and $U_0 \in SU(N_v)$. The perturbative fields $\xi$ 
satisfy
\ba
\int d^4 x~ {\rm Tr}\left[T^a \xi \right]  = 0,
\ea
where $T^a$ is a generator of the subgroup $SU(N_v)$. 

 It is convenient to include the $\theta$ dependence as
\ba
e^{i \theta} U = \left(\begin{array}{cc}
e^{i{\theta \vec{I_v} \over N_v}} U_0 & 0 \\
0   & \vec{I_s} \end{array}\right) \; \exp\left( {2 i \xi \over F}\right) =\left(\begin{array}{cc}
 {\bar U}_0 & 0 \\
0   & \vec{I_s} \end{array}\right) \; \exp\left( {2 i \xi \over F}\right) .
\ea
If the topology is fixed so that $\theta$ is integrated over,  the path integral at LO in the $\epsilon$ expansion is 
\ba
{\cal Z}_\nu & \simeq& \int d\xi~J(\xi) e^{- \int d^4 x \left(\Tr\left[\partial_\mu \xi(x) \partial_\mu \xi(x)\right] + M_{ss}^2  \Tr\left[ P_s \xi^2 \right]  \right)  } \nonumber\\
& & ~\int_{U(N_v)} ~d {\bar U}_0 \det({\bar U}_0)^\nu    e^{  {1 \over 2} \Tr\left[P_v \left(M {\bar U}_0 + {\bar U}_0^\dagger M\right)\right]},   
\ea
and the integration over the zero and non-zero modes factorizes.

The term $J(\xi)$ is the Jacobian that comes about from the change in the measure when moving from $[d U]$ to $[d {\bar U}_0] [d\xi]$, which 
will contribute at  NLO as in the $\epsilon$-regime. We describe the computation of this measure term in Appendix A. 

The integration over the $\xi$ variables is done in perturbation theory. 
In order to write the $\xi$ propagator, we need to distinguish the indices in the $v$ and $s$ sector, we denote the former by latin letters $a,b,...$ and the latter by greek ones $\alpha,\beta,...$. The propagator for the $\xi$ fields is :
\ba
\Bigl\langle \xi_{\c\a}(x) \, \xi_{\d\b}(y) \Bigr\rangle  &=& 
 \fr12 \Bigl[\delta_{\c\b} \delta_{\d\a} {\bar G}(x-y;0) - 
 \delta_{\c\a} \delta_{\d\b} \left( {\bar E}(x-y;0, 0) - {N_s \over N_v^2 V M_{ss}^2}\right )  \Bigr]\;, \label{xi_1}\\
 \Bigl\langle \xi_{\c\alpha}(x) \, \xi_{\delta\b}(y) \Bigr\rangle  &=& 
 \fr12 \delta_{\c\b} \delta_{\delta\alpha} {G}\left(x-y;{M_{ss}^2\over 2}\right)\label{xi_2}\\
  \Bigl\langle \xi_{\c\a}(x) \, \xi_{\delta\beta}(y) \Bigr\rangle &=& 
 -\fr12 
 \delta_{\c\a} \delta_{\delta\beta} {1\over N} {G}\left(x-y;  {N_v\over N} M_{ss}^2\right) \label{xi_3}\\
  \Bigl\langle \xi_{\gamma\alpha}(x) \, \xi_{\delta\beta}(y) \Bigr\rangle  &=& 
 \fr12 \Bigl[\delta_{\gamma\beta} \delta_{\delta\alpha} {G}(x-y;M_{ss}^2)  - 
 \delta_{\gamma\alpha} \delta_{\delta\beta} {\bar E}(x-y;M_{ss}^2, M^2_{ss}) \Bigr]\;, \la{xi_4}
 \ea
 where ${\bar G}(x,M^2)$ is defined in eq.~(\ref{Gx}) and 
\be
{\bar E}(x;M_{aa}^2, M_{cc}^2) \equiv 
 \frac{1}{V }
 {\sum_{n \in \zz}}   \Bigl(1 - \delta^{(4)}_{n,0} \Bigr) 
 \frac{e^{i p \cdot x}}{{(p^2 +M_{aa}^2)(p^2 + M_{cc}^2)  {F}(p)} },
 \label{xi_mixed}
 \ee
 with
 \be
 F(p) \equiv \left[ \frac{\Ns}{p^2+M_{ss}^2}+  \frac{\Nv}{p^2} \right].
  \ee

The computation of the left-current correlator at NLO, that is at relative order $\epsilon^2$ gives a result which has the same structure as in the $\epsilon$-regime
\ba
 \mathcal{C}^{mixed}(x_0) 
&=& \frac{F^2}{2 T}
 \biggl[
 1 - \frac{1}{F^2}\biggl(
  \Nv {\bar G}(0,0) + \Ns {G}\left(0,{M^2_{ss}\over 2}\right) - 8 L_4 N_s M_{ss}^2
 + \frac{T^2}{V} \left( N_v k_{00} + N_s k^s_{00} \right) \biggr) \nonumber\\
 &+& \frac{2 T^2 }{F^2 V} \mu\sigma^{(0,N_v)}_\nu(M) h_1(\hat x_0 ) 
 \biggr],  
 \; \la{Ct_mixed}
\ea
where 
\ba
\frac{T^2 k^s_{00}}{V} \equiv  T \frac{{\rm d}}{{\rm d} T} G\left(0,{M^2_{ss}\over 2}\right) .
\ea
and 
\ba
\mu\sigma^{(0,N_v)}_\nu(M) \equiv \int_{U(N_v)} ~dU ~{\mu_v \over 2 \Nv}~{\rm Tr}\left[ U + U^\dagger  \right] ~(\det U)^\nu ~\exp\left({\mu_v\over 2} {\rm Tr}\left[U+ U^\dagger \right] \right). 
\ea

\subsection{Decoupling of the $s$ quarks}

It is useful to rewrite the result of eq.~(\ref{Ct_mixed}) in a way which is almost identical to the result in the $\epsilon$-regime 
for a full theory with $N_v$ degenerate flavours but with a modified $F$:
\ba
 \mathcal{C}^{mixed}(x_0) 
 &=& \frac{{\tilde F}^2}{2 T}
 \biggl[
 1 + \frac{N_v}{F^2}\biggl(
 { \beta_1 \over \sqrt{V}}
 - \frac{T^2}{V}  k_{00}  \biggr) 
 + \frac{2 T^2 }{F^2 V} \mu\sigma^{(0,N_v)}_\nu(M) h_1(\hat x_0 ) \biggr],  \nonumber\\
 &-& \frac{N_s}{2 T} \biggl( {\bar G}_V\left(0,{M_{ss}^2 \over 2}\right) + \frac{T^2}{V} k^s_{00}\biggr)
  \;, \la{Ct_deco}
\ea
where 
\ba
{\tilde F}^2 = F^2 \biggl[  1 -  {\Ns\over F^2}  \left({G}_{\infty}\left(0,{M^2_{ss}\over 2}\right) - 8 L_4 M_{ss}^2\right) \biggr].
\label{ftilde}
\ea
The only difference between this expression and that of the full theory with $\Nv$ degenerate quarks are the finite volume effects in the second line, that are exponentially suppressed in $M_{ss} L$.

It is easy to understand these results: the $s$ quarks in the mixed-regime contribute as decoupling particles, because the mixed regime probes much lower energy scales than $M_{ss}$, since the $v$ quarks are much lighter and the size of the box is also much larger than the Compton wavelength of the
heavy pions: 
\ba
M^2_{vv} \leq L^{-2} \leq M_{ss}^2 \ll (4 \pi F)^2 . 
\ea
In this situation one can  integrate out the $m_s$ quark within the effective theory\cite{gl2,largemc,2loop}. According to general symmetry arguments we expect that the theory in this limit can be matched to a theory with $SU(N_v)$ flavour symmetry. The effects of the heavy particles can be absorbed in the low-energy couplings of the resulting effective theory. Since all the $v$ quarks are degenerate in mass, the result for the correlator should be identical to that of eq.~(\ref{Ct_eps}) with $m_s = m_v$ and $N_s +N_v \rightarrow N_v$ which is precisely what we have found, apart from exponentially suppressed finite volume effects.   In fact the result for 
the renormalized coupling, ${\tilde F}^2$, with $N_s=1$ coincides with that obtained in \cite{gl2} where  the matching of the $SU(3)$ flavour and the $SU(2)$ flavour effective theories for large strange quark mass was first considered. 

Another observation is that also within the $p$-regime we can consider a separation of scales $L^{-2} \leq M_{vv}^2 \ll M_{ss}^2$.  A similar factorization would then be possible for correlators involving only $v$ quarks as external legs, and up to exponentially suppressed terms in $M_{ss} L$. The result can be written
as the correlator in the $p$-regime for $N_v$ degenerate quarks with mass $m_v$ with modified couplings $\tilde F$ as in eq.~(\ref{ftilde}) and $\tilde \Sigma$:
\ba
\tilde{\Sigma} = \Sigma \left( 1 - {\Ns\over F^2} {G}_{\infty}\left(0,{M^2_{ss} \over 2}\right) + \frac{E^{sub}_{\infty}(0,0,0)}{F^2} +{16 \over F^2} \Ns M_{ss}^2   L_6 \right),
\ea
where 
\be
E^{sub}_{\infty}(x,M_{vv}^2, M_{vv}^2)  \equiv  E_{\infty} (x,M_{vv}^2 , M_{vv}^2) -  {1\over N_v} G_{\infty} (x,M_{vv}^2), 
  \label{ep}
 \ee
which also coincides with the result of \cite{gl2}. 

\begin{figure}[htbp]
\begin{center}
\includegraphics[width=8cm]{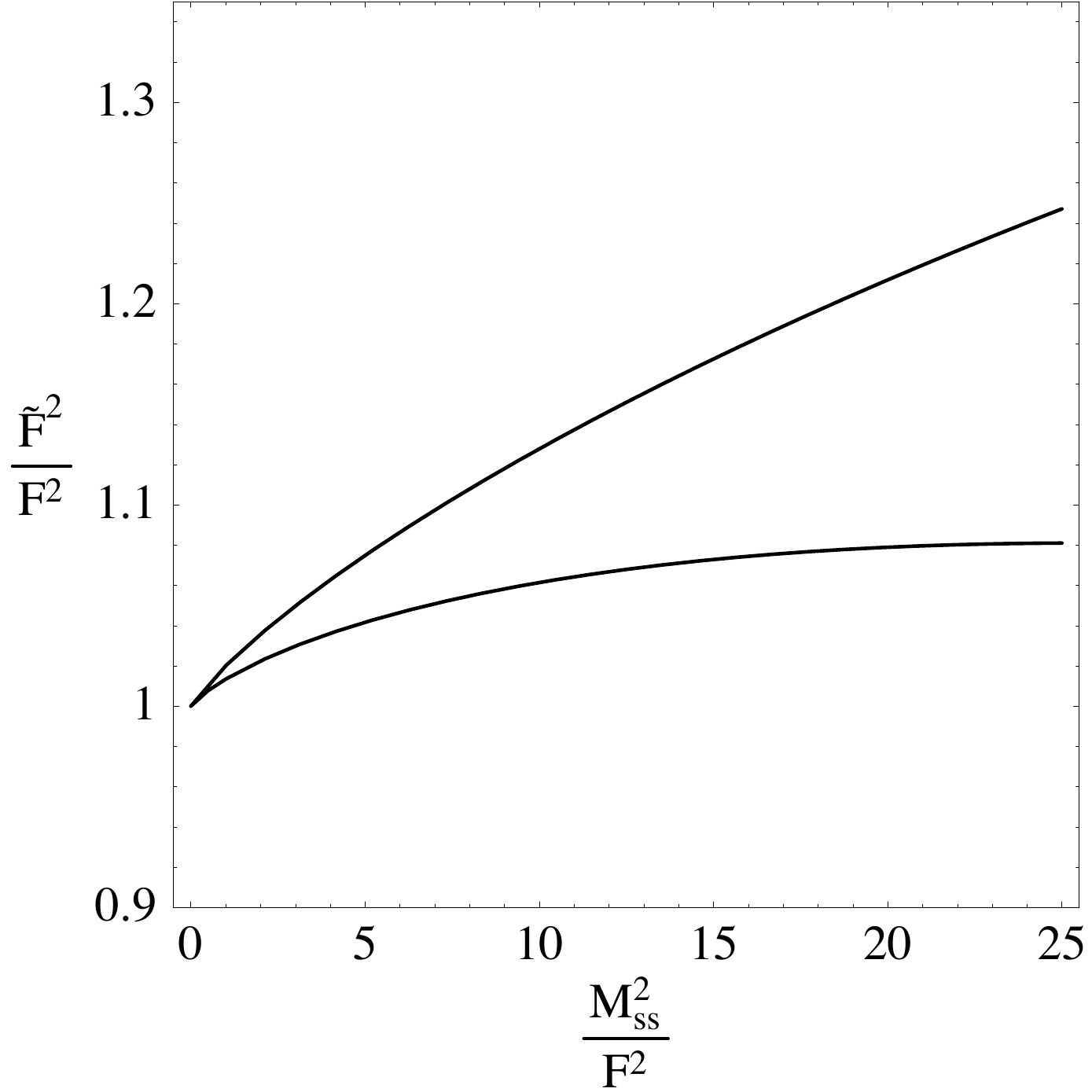}\includegraphics[width=8cm]{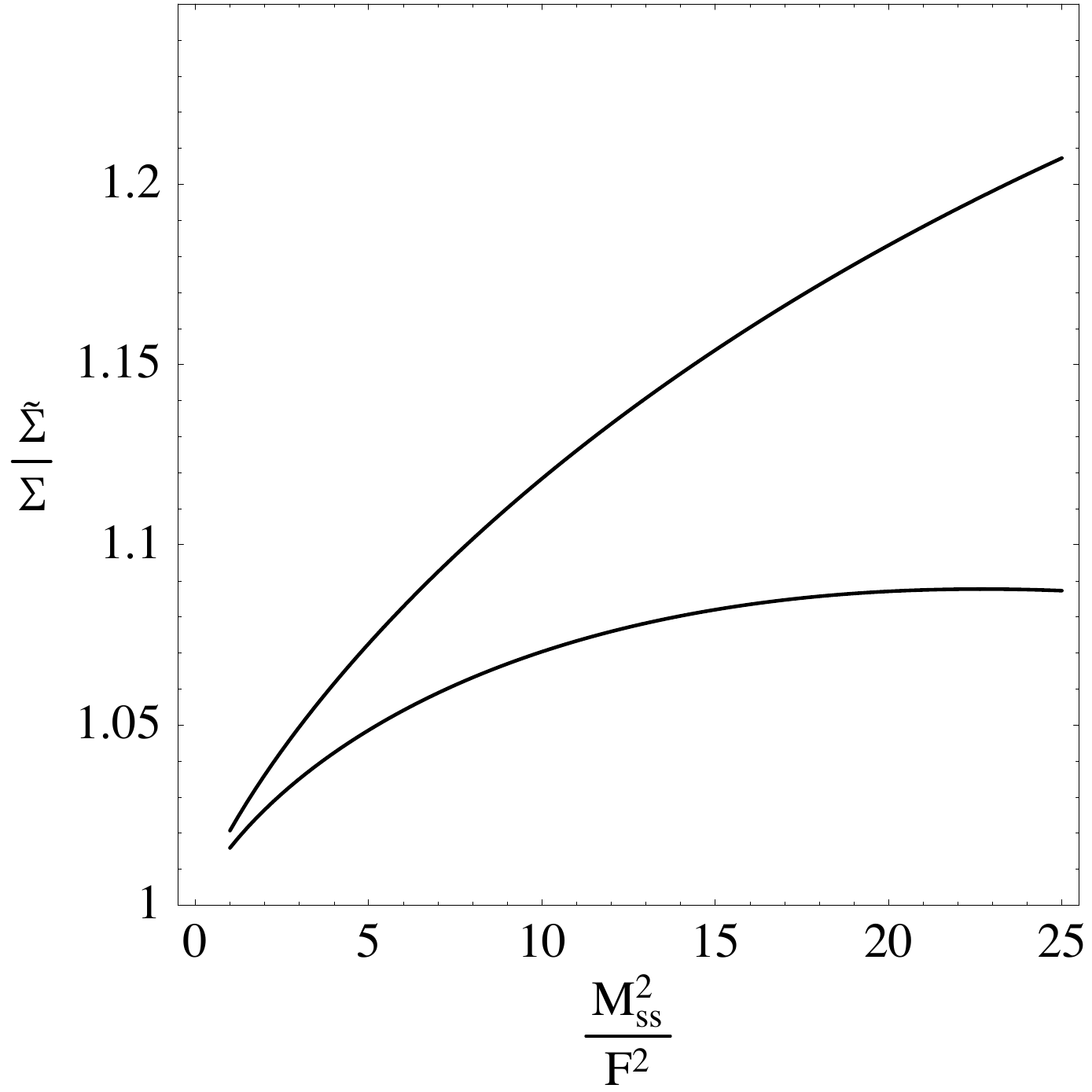}
\caption{ Dependence of the effective couplings ${\tilde F}^2$ (left) and $\tilde\Sigma$ (right) in a theory with $N_v=2$, and with $N_s=1$ quarks integrated out, as functions of $M_{ss}^2/F^2$ for $F=90$~MeV. The two lines correspond to the extreme values of $L_4, L_6$ obtained  
in the phenomenological determinations reviewed in  \cite{ecker}. }
\label{fig:tilde}
\end{center}
\end{figure}

In Figures~\ref{fig:tilde} we show the $\tilde F$ and $\tilde \Sigma$ as functions of $M_{ss}^2 /F^2$.  \vspace{0.5cm}

 The reason that $\tilde \Sigma$  does not appear in the mixed-regime of eq.~(\ref{Ct_mixed}) is because $\Sigma$ appears there only at NLO and therefore any correction to it, would be of  higher order. 

It is important to stress however that the decoupling only works in a finite volume up to exponentially suppressed 
corrections in $M_{ss} L$, since there is no way to predict these finite volume corrections within the 
effective theory after the $s$ quarks are integrated out.

\section{Partially-quenched Theory}
\label{sec:pq}

We will know consider a partially-quenched theory in which there are $\Nv$ quenched valence quarks of mass $m_v$ and $\Ns$ sea quarks of mass $m_s$. Note that we consider the generators appearing in the left-currents belong to the valence subgroup. 

In order to obtain the partially-quenched results \cite{pq}, it is simpler to use 
 the replica method of \cite{replica}. In this method one enlarges the valence sector  to $N_r$ degenerate flavours of mass $m_v$, where  only $\Nv$ of these appear in the external sources. The full symmetry group for zero quark masses is therefore $SU(N_r + \Ns)$ and the path integral of this theory at the quark level is 
 \ba
{ \cal Z}[J] = \int \left[d A_\mu\right] ~\det\left(\not\!\! D + m_v + J \right)^{N_v}~\det\left(\not\!\! D + m_v \right)^{N_r-N_v}~\det\left(\not\!\! D + m_s \right)^{N_s}~e^{-S_g\left[A_\mu\right]}.
 \ea 
 Taking the limit $N_r \rightarrow 0$ in this expression, one obtains the supersymmetric formulation of the 
 theory \cite{ddhj}. The replica method therefore dictates that one should construct the chiral effective theory 
 for the $N=N_r+N_s$ flavours, do perturbative calculations keeping the explicit dependence on $N_r$  and take the limit 
 $N_r \rightarrow 0$ at the end. We know consider the three regimes in this context.

\subsection{$p$-regime}

In the $p$-regime, 
 as long as $\Ns \neq 0$, the replica limit can be taken and coincides with the result obtained using the supersymmetric method \cite{pq}.  The $\xi$ propagator is that of eq.~(\ref{gen_prop}) with $N_v \rightarrow N_r$. 
  
 The result for the left-current correlator is that of eq.~(\ref{Ct_p}) with $N_v \rightarrow N_r = 0$ and 
 \ba
\Delta^{Pq}_F &=&  - \frac{\Ns}{F^2} G(0;M_{vs}^2)  + \frac{8}{F^2}  \left(\Ns M_{ss}^2 L_4+ M_{vv}^2 L_5 \right), \\
\Delta^{Pq}_M  &=&  \left[\frac{E(0;M_{vv}^2,M_{vv}^2)}{F^2} \right]_{N_v=0}- \frac{8}{F^2} 
\Bigl( \Ns M_{ss}^2 (L_4 -2 L_6) + M_{vv}^2 (L_5 - 2 L_8)\Bigr).
\ea
In the  limit  $m_s \rightarrow m_v$, the full theory result for $N_s$ degenerate flavours is recovered. 
Results for the meson masses and decay constants at  NNLO have been recently obtained \cite{bdl}.

It is important to realize that in the partially-quenched theory, the full set of ${\cal O}(p^4)$ couplings
need to be used. The reduction of independent couplings in the full theory with $N=2$ or $N=3$ only takes place in the unquenched limit, ie. $m_v \rightarrow m_s$. In particular this implies that for 
$N_s=2$, the partially-quenched predictions involve more couplings that those that are physical in the unquenched limit. Obviously these couplings cannot be determined from phenomenology (not even in principle) and need to be determined on the lattice.

\begin{figure}[htbp]
\begin{center}
\includegraphics[width=8cm]{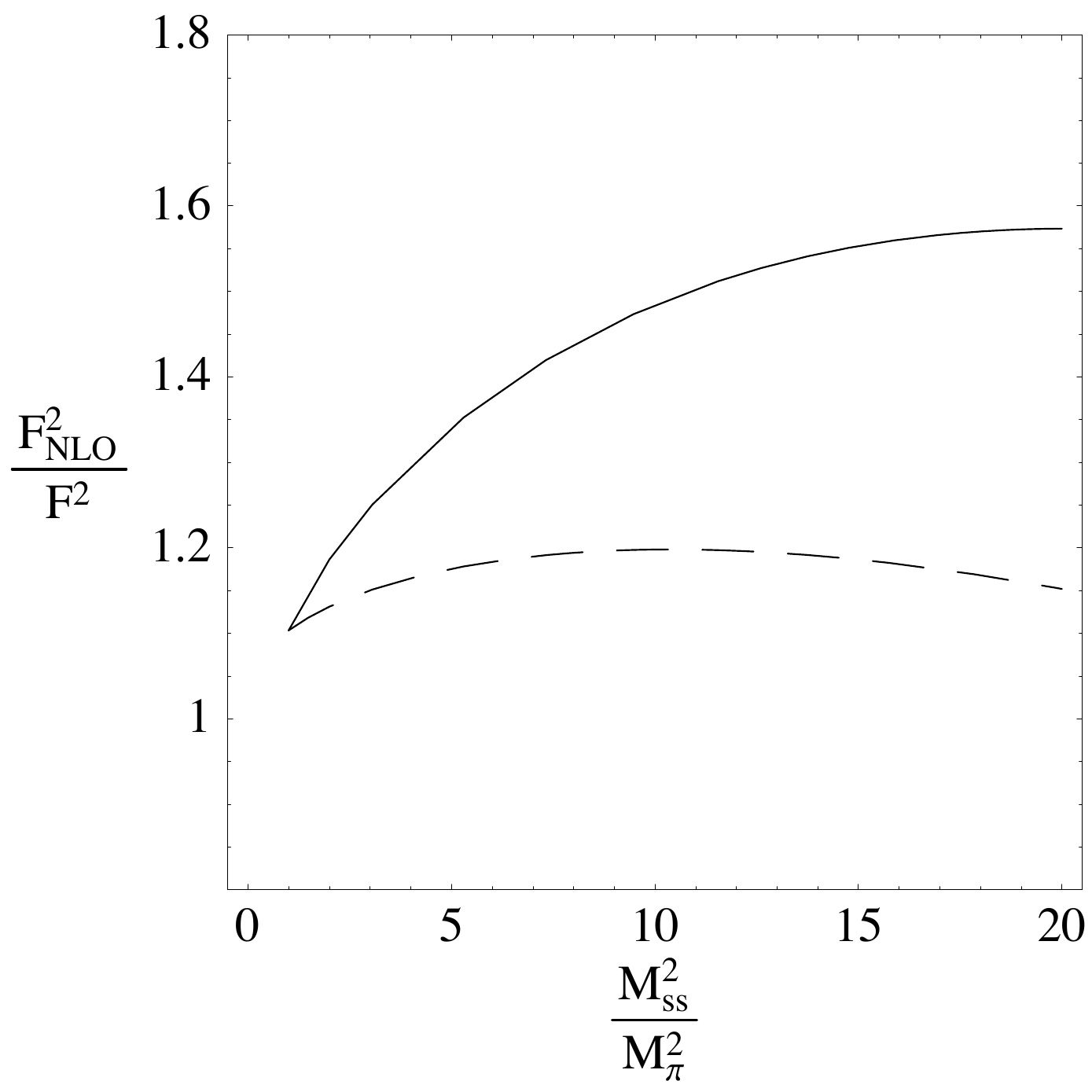}\includegraphics[width=8cm]{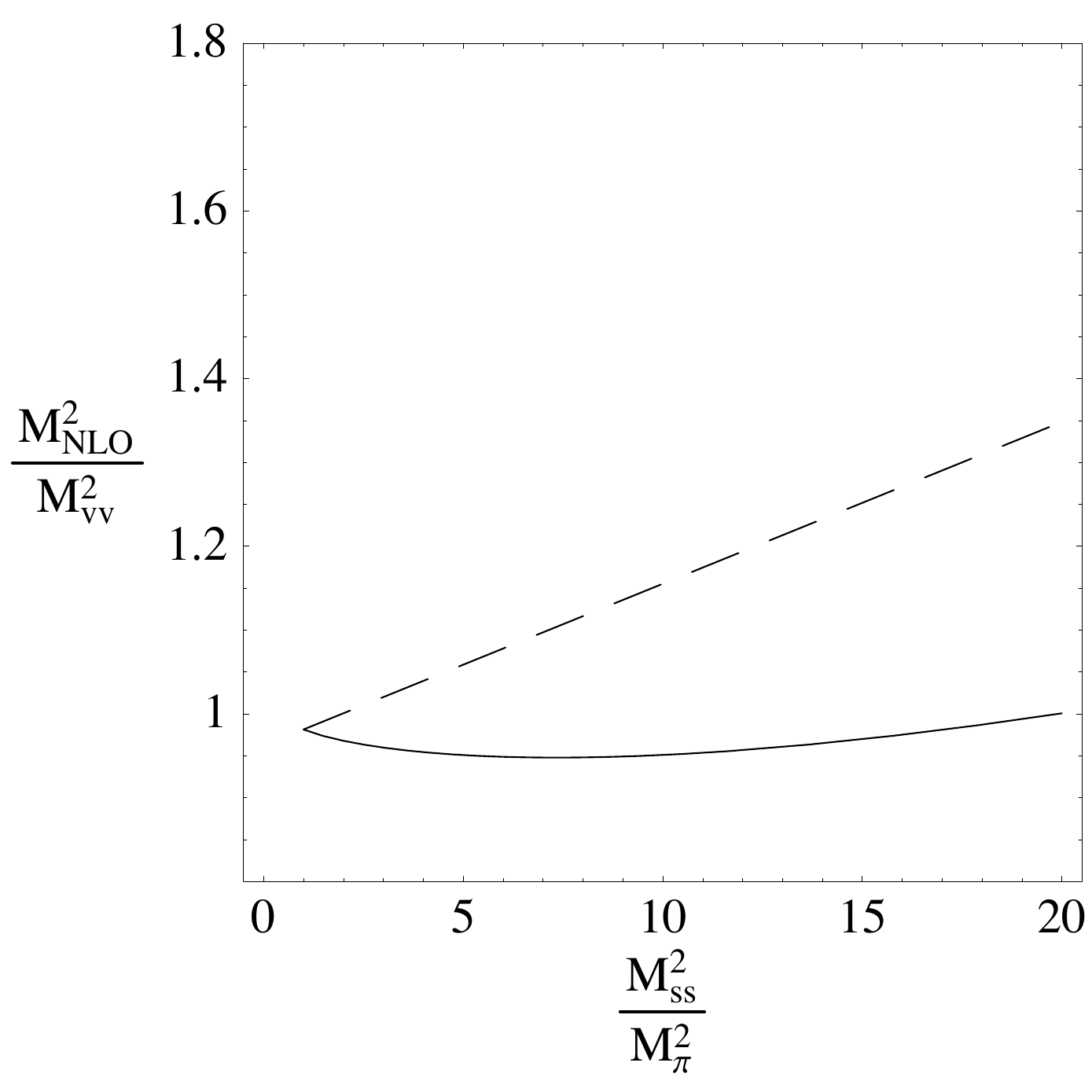}
\caption{ Chiral correction of $F^2$ (left) and $\Sigma$ (right) as a function of $m_s$, with $M_{vv}= M_\pi= 135 $MeV in the partially-quenched $N_s=2$ theory (dashed line), or of $m_s = m_v$ in the full case with $N_s+N_v = 2$ (solid line), in a lattice of 5~fm. The values of the low-energy couplings have been chosen as the central values in the phenomenological determinations of \cite{gl2,ecker}:  $L_4(M_\rho)=-0.3\cdot 10^{-3}$, $L_5(M_\rho)=1.4\cdot 10^{-3}$,$L_8(M_\rho)=0.9\cdot 10^{-3}$ and $L_6(M_\rho)=-0.2\cdot 10^{-3}$ with $F=90$~MeV.}
\label{fig:p1}
\end{center}
\end{figure}

An interesting observation is that the partially-quenched correction to the meson mass $\Delta^{Pq}_M$ has no 
logarithm in the sea-quark mass, just in the valence quark. If valence quarks masses could be simulated in the light regime, for example using Ginsparg-Wilson fermion regularizations \cite{gw}, and only the sea quark masses would be kept unphysically large, the $m_s$ dependence of the meson mass  would be strictly linear at this order of the chiral expansion. In the case of the decay constant the logarithm remains 
but with a smaller coefficient. These features are shown in Figure \ref{fig:p1} where we show the dependence of the meson mass and decay constant with the sea-quark mass for a value of the valence quark mass of $5~$MeV for $N_s=2$.  This is compared with the $m_s = m_v$ dependence in the full theory case, for $N_s+N_v =2$.  

\vspace{0.5cm}

In order to recover the fully-quenched case  $N_s \rightarrow 0$, it is necessary to keep the singlet 
meson in the theory. When the singlet with a mass $m_0^2$ is kept in the theory  the 
singlet part of the propagator in eq.~(\ref{e}) is modified to 
 \ba
 E^q(x; M_{aa}^2, M_{cc}^2) \equiv 
 \frac{1}{V} 
 \sum_{n \in \zz} 
 \frac{e^{i p \cdot x} \left({\alpha p^2 + m_0^2 \over 2 N_c}\right)}{{(p^2+M_{aa}^2)(p^2+M_{cc}^2){F^q}(p)} },
 \ea
 with 
 \ba
 F^q(p) \equiv  1+ \left({\alpha p^2 + m_0^2\over 2 N_c}\right)  \left[ \frac{\Ns}{p^2+M_{ss}^2}+  \frac{N_r}{p^2+M_{vv}^2} \right] ,
  \ea
  which is well-defined for $\Ns=N_r=0$. 
Note that as long as either $\Ns$ or $N_r$ are different from zero, the limit $m_0^2 \rightarrow \infty$ can be safely taken. The results for the two-point function in this limit agree with those obtained in \cite{weak}. 

\subsection{$\epsilon$-regime}

The results for the partially-quenched theory well all quarks are in the $\epsilon$-regime are 
\be
 \mathcal{C}^{\epsilon,Pq}(x_0) 
 = \frac{F^2}{2 T}
 \biggl[
 1 + \frac{N_s}{F^2}\biggl(
 \frac{\beta_1}{\sqrt{V}} - \frac{T^2 k_{00}}{V} \biggr)
 + \frac{2 T^2 }{F^2 V} \mu\sigma^{Pq(N_s,N_v)}_\nu(M) h_1(\hat x_0 ) 
 \biggr], 
 \;, \la{Ct_eps}
\ee
where 
\ba
\mu\sigma^{Pq(N_s,N_v)}_\nu(M)  \equiv \lim_{N_r \rightarrow 0} \int_{U(N_s+N_r)} ~dU ~{\mu_v \over 2 \Nv}~{\rm Tr}\left[ \Pv U + U^\dagger  \Pv\right] ~(\det U)^\nu ~\exp\left({\Sigma V \over 2} {\rm Tr}\left[ M U + U^\dagger M \right] \right), \nonumber\\
%\label{musigmapq}
\ea
where $P_v$ is the projector on the $SU(N_v)$ subgroup of $SU(N_r)$. 
 This limit has been defined in \cite{znupq} as 
\ba
\mu\sigma^{Pq(N_s,N_v)}_\nu(M) \equiv \int_{Gl(\Ns+N_v|N_v)} ~\!\!\!\!\!dU ~{\mu_v \over 2 N_v}~{\rm Tr}\left[P_v U + U^\dagger  P_v \right] ~(\det U)^\nu ~\exp\left({\Sigma V \over 2} {\rm Tr}\left[ M U + U^\dagger M \right] \right), \nonumber\\
\label{musigmapq}
\ea
where $U$ is an element of the maximal Riemannian manifold, $Gl(N_s+N_v |N_v)$. This integral can be obtained as a functional derivative of the functional
\ba
Z^{Pq(\Ns,N_v)}_\nu(M_J) = \frac{\det\left[  \mu_i^{j-1} {\cal I}_{\nu+j-1}(\mu_i)\right]_{i,j=1,...,N_v}}{
\prod_{j > i=1,..,N_v} (\mu_j^2 - \mu_i^2) \prod_{j > i=N_v+1,..,N} (\mu_j^2 - \mu_i^2)},
\label{znupq}
\ea
with 
\ba
{\cal I}_\nu(x_i) = \left\{\begin{array}{cl}
(-1)^\nu K_\nu(x_i) \; & i=1,..,N_v \\
I_\nu(x_i)\;  & i=N_v+1,...,2 N_v+N_s \\
\end{array} \right.
\ea
$\mu_i = \mu_v , i=1,..,N_v; \mu_i=(m_v+J)\Sigma V, i=N_v+1,...,2 N_v$ and $\mu_i= \mu_s \equiv m_s \Sigma V, i=2 N_v +1,...,2 N_v + N_s$.

$Z^{Pq(N_s,N_v)}_\nu(M)$ is the same for any value of $N_v$. This is easy to check for small values of $N_v$ and $N_s$. Essentially the choice of $N_v$ is determined by the dimension of the 
external source, $N_v$ in eq.~(\ref{znupq}). In order to obtain the function $\mu\sigma^{Pq(N_s,N_v)}_\nu(M)$ it is enough  to have an external source coupled to one of the valence quarks, since due to the $SU(N_v)$ invariance, the contribution of each $v$ quark to eq.~(\ref{musigmapq}) is the same. We can therefore choose the simplest case, $N_v=1$. Any other choice would give the same result.
 
 Let us consider two simple examples. 
 
 \vspace{0.25cm}
 
 {\it Example 1}: $N_v=1, N_s=1$

It is easy to check that the partition functional is that of a one flavour theory:
\ba
Z_\nu^{Pq(1,1)}(M) = I_\nu(\mu_s) ,
\ea
while the function $\mu\sigma^{Pq(1,1)}_\nu(M)$ is:
\ba
\mu\sigma_\nu^{Pq(1,1)}(M) = \mu_v \Sigma^{q}(\mu_v) + \frac{2 \mu_v^2}{\mu_v^2 - \mu_s^2} \frac{K_\nu(\mu_s)}{I_\nu(\mu_s)} \left[ \mu_s I_{\nu+1}(\mu_s) I_\nu(\mu_v) - \mu I_{\nu+1}(\mu_v) I_\nu(\mu_s)\right] , 
\ea
where $\Sigma^{q}(\mu)$ is the quenched quark condensate in the $\epsilon$-regime at LO\cite{dotv}:
\ba
\Sigma^{q}(\mu) \equiv   \mu \left[ I_\nu(\mu) K_\nu(\mu) + I_{\nu+1}(\mu) K_{\nu-1}(\mu) \right] + {\nu \over \mu}.
\ea 
It is easy to check from this expression that the quenched limit is obtained as $\mu_s \rightarrow \infty$:
\ba
\lim_{\mu_s \rightarrow \infty}  \mu\sigma_\nu^{Pq(1,1)}(M) = \mu_v \Sigma^{q}(\mu_v),  
\ea
and the full theory with just one flavour $N_s=1$ is obtained when the valence and sea masses are the same, that is in the limit $\mu_s \rightarrow \mu_v$:
\ba
\lim_{\mu_s \rightarrow \mu_v}  \mu\sigma_\nu^{Pq(1,1)}(M) = \mu\sigma^{(0,1)}(\mu_v) = \mu_v {I'_\nu(\mu_v) \over I_\nu(\mu_v)} . 
\ea
Note that this last result is a {\it non-perturbative} test of Sharpe and Shoresh conjecture that the full theory with $N_s$ flavours can be smoothly obtained from the partially-quenched approximation with $\Ns$ sea and $\Nv$ valence quarks.
 
 \vspace{0.25cm}
 
 {\it Example 2}: $N_v=1$, $N_s=2$
 
 The functional for this case if the one corresponding to a two-flavour theory: 
 \ba
 Z_\nu^{Pq(2,1)}(M) = {1\over 2} \left[ I_\nu(\mu_s)^2 - I_{\nu+1}(\mu_s) I_{\nu-1}(\mu_s)\right] ,
 \ea
 while the function $\mu\sigma^{Pq(2,1)}_\nu(M)$:
 \ba
\mu\sigma_\nu^{Pq(2,1)}(M) = \mu_v \Sigma^{q}(\mu_v) - \frac{2 \mu_v^2}{\mu_v^2 - \mu_s^2} + {2 \mu_v^3 \mu_s \over (\mu_v^2 -\mu_s^2)^2} \frac{G_\nu(\mu_s,\mu_v)}{Z_\nu^{Pq(2,1)}(\mu_s)},
\ea
where 
\ba
G_\nu(\mu_s,\mu_v) &\equiv& I_\nu(\mu_s) I_{\nu+1}(\mu_s) ( I_{\nu+1}(\mu_v) K_\nu(\mu_v) - I_\nu(\mu_v) K_{\nu+1}(\mu_v)) \nonumber\\
&+& {\mu_v\over \mu_s} I_\nu(\mu_s)^2 K_{\nu+1}(\mu_v) I_{\nu+1}(\mu_v) -  {\mu_s\over \mu_v}  I_{\nu+1}(\mu_s)^2 K_{\nu}(\mu_v) I_{\nu}(\mu_v).
\ea
As in the previous example one can explicitely check that in the limit $\mu_s \rightarrow \infty$ one recovers the quenched limit, while in the limit $\mu_s \rightarrow \mu_v$ one recovers the full theory with $N=2$ degenerate quarks.

\begin{figure}[htbp]
\begin{center}
\includegraphics[width=8cm]{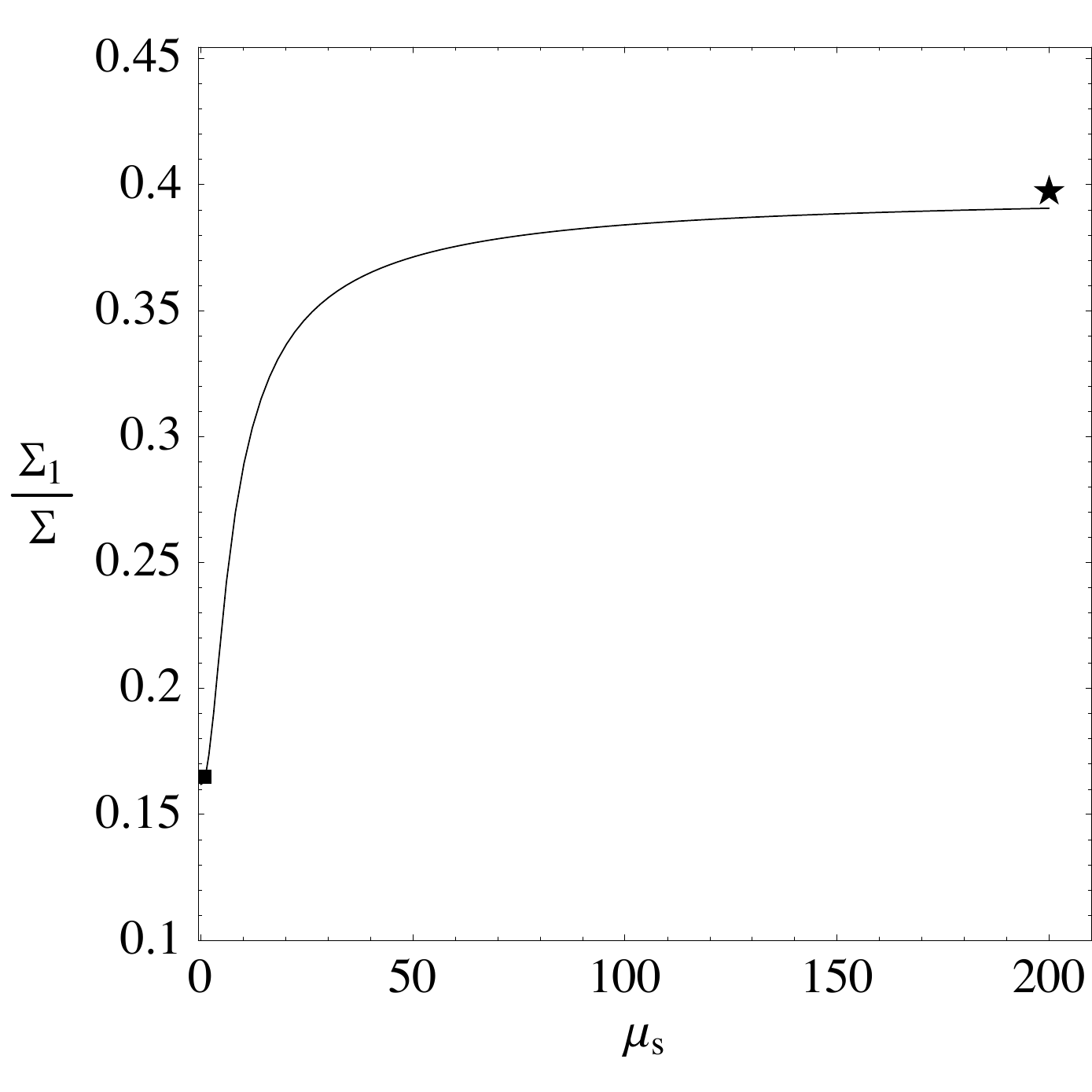}\includegraphics[width=8cm]{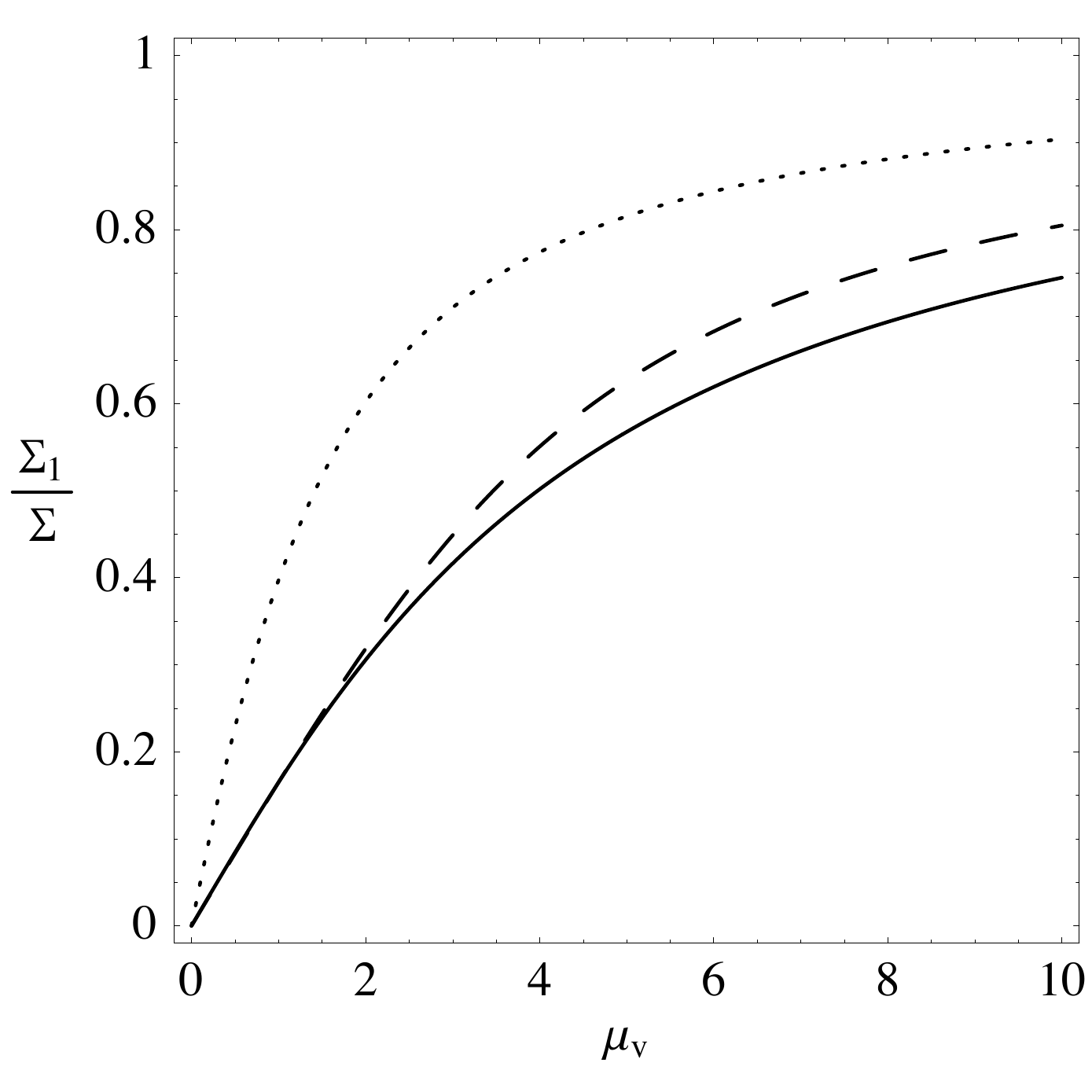}
\caption{Left: Function $(\mu\sigma^{Pq(2,1)}_\nu(M)-|\nu|)/\mu_v$ as a function of $\mu_s\equiv m_s \Sigma V$ for $\mu_v\equiv m_v \Sigma V = 1$. The star and the dot correspond to the quenched and the full $(0,2)$ result respectively at the same $\mu_v$. Right: The same function as a function of $\mu_v$ for $\mu_s=1$ for topology $\nu=1$ (solid), compared with the quenched function (dotted) and the full $(0,2)$ (dashed). }
\label{fig:cond_e}
\end{center}
\end{figure}

The partially-quenched result then interpolates between the quenched and the full theory with $N_s$ flavours. This is shown in left figure of Figure \ref{fig:cond_e} where  the function 
$(\mu\sigma^{Pq(2,1)}_\nu(M)-|\nu|)/\mu_v$ is shown for the partially quenched case as a function of $\mu_s$ and compared with 
full $N=2$ ($\mu_s = \mu_v$)  and quenched results. On the right figure the $\mu_v$ dependence of the condensate 
for two topological sectors is shown and compared with the quenched and full theory, setting $\mu_s=1$.

\vspace{0.5cm}

\subsection{Mixed-regime}

Both in the $p$ and $\epsilon$ regimes we could obtained the partially-quenched result from the full one with $N_s+N_r$ quarks by taking the limit $N_r \rightarrow 0$ at the end of the calculation. While for the current correlator the limit seems to be well-defined and indeed is the right answer, in other correlation functions such as the pseudoscalar correlator, the limit does not exist. It is easy to see this simply by looking at the $\xi$ propagators of eq.~(\ref{xi_1})-(\ref{xi_4}). In the partially-quenched theory the $p$-regime propagator is the same but with $N_v \rightarrow N_r$. If we try to take the $N_r \rightarrow 0$ limit, the zero-momentum mode contributions in the first and third equations  explode. This is 
exactly the same effect that happens in the quenched case if the singlet field is not kept in the effective theory \cite{pq}. 

In the partially -quenched case, the $U(N_s+N_r)$ singlet can be integrated out and this is true no matter whether we are in the $p$, $\epsilon$ or mixed regimes. However, what plays the role here of the non-decoupling singlet is the traceless generator of the flavour group $SU(N_s + N_r)$, that is a singlet under the $SU(N_r)$ subgroup, whose normalized generator is
\ba
 T_{\eta}=\sqrt{\frac{N_r N_s}{2 (N_s+N_r)}}diag\{\underbrace{\frac{1}{N_r},\dots,\frac{1}{N_r}}_{N_r},\underbrace{-\frac{1}{N_s},\dots ,-\frac{1}{N_s} }_{N_s}\}\quad .
\ea
In the mixed-regime not only the $SU(N_r)$ generators become massless in the limit $m_v \rightarrow 0$, in the limit $N_r \rightarrow 0$, also the pion associated with the $\eta$ field 
 gets massless. In fact the propagator for this field can be easily derived from eq.~(\ref{xi_mixed}) 
\ba
\Bigl\langle \eta(x) \, \eta(y) \Bigr\rangle  &=& 
 \fr12 {G}\left(x-y;{N_r \over N} M_{ss}^2\right) , 
\ea
and therefore becomes massless if $N_r \rightarrow 0$ and its zero-mode contribution diverges. Note that in the full case this is however a massive mode, since the mass goes with the largest massgap.

The way out of this problem is to modify the factorization in such a way that the zero-momentum mode of the $\eta$ field is also treated non-perturbatively. So instead of eq.~(\ref{factor_mixed}) we have
\ba
U = \left(\begin{array}{cc}
e^{i {\eta \over N_r} \vec{I_r} } U_0   & 0 \\
0   & e^{-i {\eta \over N_s} \vec{I_s}}   \end{array}\right) \; \exp\left( {2 i \xi \over F}\right),
\ea
and now the $\xi$ field do not contain the zero-modes of the $SU(N_r)$ generators nor  the $\eta$ one. 
As in the quenched theory \cite{ddhj}, the LO Lagrangian would not factorize into $\xi$ and $U_0$ fields
in this case, however it does after the integration over $\theta$ is performed, that is in a fixed topological sector. In Appendix B we derive the path integral and show that at the LO in the $\epsilon$ expansion, the integration over the zero and non-zero modes factorize: 
\ba
{\cal Z}^{LO}_\nu = \int_{U(N_r)} d {\bar U}_0  ~ e^{{\mu_v \over 2} \Tr\left[{\bar U}_0+{\bar U}_0^\dagger\right] } ~\int d\xi e^{- \int d^4 x ~\Tr\left[\partial_\mu \xi(x) \partial_\mu \xi(x)\right] + {M_{ss}^2} \Tr\left[ P_s \xi^2\right]  }. 
\ea

The new $\xi$ propagator is:
\ba
\Bigl\langle \xi_{\c\a}(x) \, \xi_{\d\b}(y) \Bigr\rangle  &=& 
 \fr12 \Bigl[\delta_{\c\b} \delta_{\d\a} {\bar G}(x-y;0) - 
 \delta_{\c\a} \delta_{\d\b}  {\bar E}^{pq}(x-y;0, 0)   \Bigr]\;, \la{gen_prop_pq_1}\\
 \Bigl\langle \xi_{\c\alpha}(x) \, \xi_{\delta\b}(y) \Bigr\rangle  &=& 
 \fr12 \delta_{\c\b} \delta_{\delta\alpha} {G}\left(x-y;{M_{ss}^2\over 2}\right) \la{gen_prop_pq_2}\\
  \Bigl\langle \xi_{\c\a}(x) \, \xi_{\delta\beta}(y) \Bigr\rangle &=& 
 -\fr12 
 \delta_{\c\a} \delta_{\delta\beta} {1\over N} {{\bar G}}\left(x-y;  {N_r\over N} M_{ss}^2\right)\la{gen_prop_pq_3}\\
  \Bigl\langle \xi_{\gamma\alpha}(x) \, \xi_{\delta\beta}(y) \Bigr\rangle  &=& 
 \fr12 \Bigl[  \delta_{\gamma\beta} \delta_{\delta\alpha} {G}(x-y;M_{ss}^2)  - \nonumber\\ 
 & & \delta_{\gamma\alpha} \delta_{\delta\beta} \left(\bar{E}^{pq}(x-y;M_{ss}^2,M_{ss}^2) +{1 \over N_s V M_{ss}^2}\right)\Bigr]\; , \la{gen_prop_pq_4}
 \ea
 where the latin indices refer to the valence and the greek to the sea, and 
 \be
{\bar E}^{pq}(x;M_{aa}^2, M_{cc}^2) \equiv 
 \frac{1}{V }
 {\sum_{n \in \zz}}   \Bigl(1 - \delta^{(4)}_{n,0} \Bigr) 
 \frac{e^{i p \cdot x}}{{(p^2 +M_{aa}^2)(p^2 + M_{cc}^2)  {F}^{pq}(p)} },
 \label{xi_mixed}
 \ee
with
 \be
 F^{pq}(p) \equiv \left[ \frac{\Ns}{p^2+M_{ss}^2}+  \frac{N_r}{p^2} \right].
  \ee

It is easy to check that the propagator is now well-defined in the limit $N_r \rightarrow 0$. The singlet part of the propagator of the $v$ modes in eq.~(\ref{gen_prop_pq_1}) has a double pole structure as in the quenched case, but instead of the singlet mass, what appears in the numerator is the heavy mass gap, $M_{ss}^2$. This double pole is a non-decoupling effect that only appears because the theory is partially-quenched.

With this parametrization it is easy to check that the left-current propagator is 
\ba
 \mathcal{C}^{mixed,Pq}(x_0) 
&=& \frac{F^2}{2 T}
 \biggl[
 1 - \frac{N_s}{F^2}\biggl(
  {G}\left(0,{M^2_{ss}\over 2}\right) - 8 L_4 M_{ss}^2
 + \frac{ T^2}{V}  k_{00}^s  \biggr) \nonumber\\
 &+& \frac{2 T^2 }{F^2 V} \mu\sigma^{Pq(0,N_v)}_\nu(M) h_1(\hat x_0 ) 
 \biggr],
 \; \la{Ct_mixed_pq}
\ea
where $\mu\sigma^{Pq(0,N_v)}(M)$ is also the fully quenched result.
  No double-pole appears for the same reason that it did not appear in the quenched case: this observable is not sensitive to them at NLO. 

\begin{figure}[htbp]
\begin{center}
\includegraphics[width=8cm]{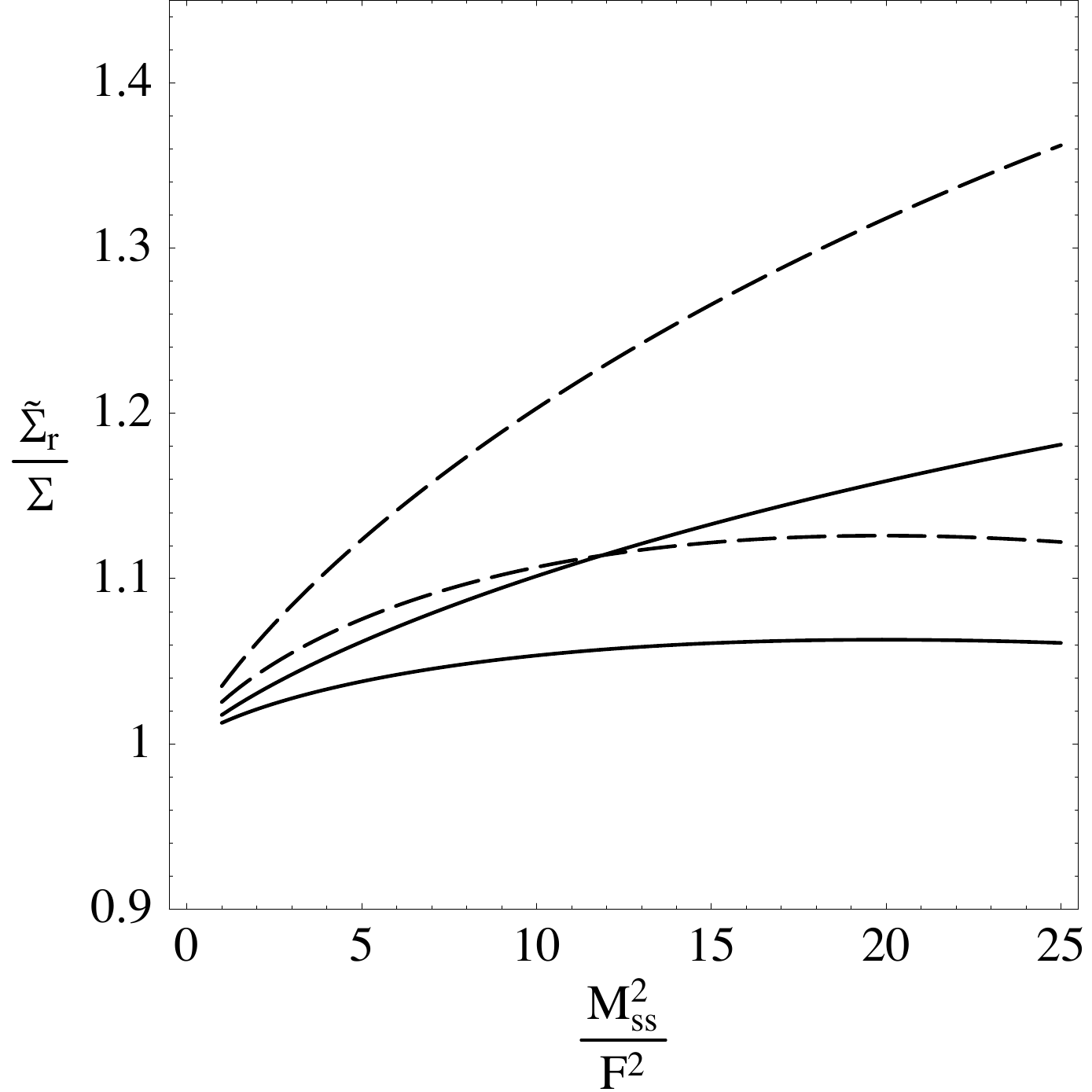}
\caption{ Dependence of the effective couplings $\tilde\Sigma_r$ on $M_{ss}^2/F^2$ for the partially-quenched theory with $N_s=1$ (solid) and $N_s=2$ (dashed) . $F=90$~MeV and the two lines correspond to the two extreme values of $L_6$ obtained  
in the phenomenological determinations reviewed in  \cite{ecker}. Of course in reality $L_6$ can be different in both cases.}
\label{fig:tildepq}
\end{center}
\end{figure}

The decoupling of the $s$ quarks is not possible in the partially-quenched case, because the $\eta$ field remains light. However, we expect that we should be able to integrate out 
the scale associated to $M_{ss}$  and match the result to a quenched effective theory.  Provided $M_{ss} \ll 4 \pi F$, this integration can be done perturbatively.  The quenched Chiral Lagrangian contains additional couplings besides $F$ and $\Sigma$: $m_0^2$ and $\alpha$ in the standard notation (a mass of the $\eta$ field and a kinetic term). The tree-level matching of 
$m_0^2$ and $\alpha$ can be easily read from the propagator of eq.~(\ref{gen_prop}). The expected $p$-regime propagator in a quenched theory with $\Nv$ valence quarks would be of the form 
\ba
\Bigl\langle \xi_{\c\a}(x) \, \xi_{\d\b}(y) \Bigr\rangle = \fr12 \Bigl[\delta_{\c\b} \delta_{\d\a} { \tilde G}(x-y;M_{ab}^2) - 
 \delta_{\c\a} \delta_{\d\b}  {\tilde E}(x-y;M_{aa}^2 , M_{cc}^2)   \Bigr]\;, 
\ea 
with $\tilde {G} =  G$,  and
\ba
\tilde  { E}(x; M_{aa}^2, M_{cc}^2) \equiv 
 \frac{1}{V} 
 \sum_{n \in \zz} 
 \frac{e^{i p \cdot x} \left({\alpha p^2 + m_0^2 \over 2 N_c}\right)}{(p^2+M_{aa}^2)(p^2+M_{cc}^2) }. 
 \ea
Identifying $\tilde {E}$ with $E$ in eq.~(\ref{e}) with $N_v \rightarrow N_r = 0$ we find
\ba
{\alpha \over 2 N_c} = {1 \over \Ns} ,\;\;\;\; {m_0^2\over 2 N_c} = {M_{ss}^2\over \Ns} . 
\label{momatch}
\ea
The quenched  $\tilde F$ is still the same as that of eq.~(\ref{ftilde}), but the matching of the quenched $\tilde \Sigma$ gives instead
\ba
\tilde{\Sigma} \left( 1 + \frac{E_{\infty}(0,0,0)}{F^2} \right) = \Sigma \left( 1 - {\Ns\over F^2} {G}_{\infty}\left(0,{M^2_{ss} \over 2}\right) + \frac{E_{\infty}(0,0,0)}{F^2} +{16 \over F^2} \Ns M_{ss}^2   L_6 \right).
\label{tildepq}
\ea
$\tilde\Sigma$ gets renormalized in the $m_v \rightarrow 0$ limit in the quenched theory, as is well-known. Therefore at one-loop the renormalized coupling is
\ba
{\tilde \Sigma}_r = \tilde{\Sigma} \left( 1 + \frac{E^{UV}_{\infty}(0,0,0)}{F^2} \right) ,
\ea
while the logs in $E_\infty$ would cancel on the two sides of the eq.~(\ref{tildepq}). The curve ${\tilde \Sigma}_r/\Sigma$ as a function of $M_{ss}^2/F^2$ is also shown for $N_s=1, 2$ in  Figure~\ref{fig:tildepq}.

\section{Conclusions}
\label{sec:conclu}

We have computed the current correlator at next-to-leading order in finite volume Chiral Peturbation Theory for non-degenerate quark masses in several interesting regimes. The case when all the quarks 
are in the $p$-regime could have been easily obtained from earlier literature \cite{gl2,pq}. However 
we have also considered the case when all quarks are in the $\epsilon$-regime both in the full and partially-quenched theories.  In the latter case our results provide a non-perturbative check of the
Sharpe-Shoresh conjecture, concerning the possibility to recover QCD from partially-quenched  approximations. 

Finally we have also  considered a regime in which some of the quarks masses are in the $\epsilon$ and some in the $p$. This mixed-regime required a new zero versus non-zero mode factorization, that we introduced and worked out in detail. A further modification was required to treat the partially-quenched theory in this case. Our results in the mixed-regime show that the quarks in the $p$-regime behave essentially as decoupling particles, so that the correlator (up to some exponentially suppressed 
finite volume corrections) is that of the $\epsilon$-regime for a theory with a reduced number of flavours (i.e. those in the $\epsilon$-regime), but with corrected low-energy couplings by the heavier quarks (i.e. those in the $p$-regime).

These results can be useful for matching lattice QCD and Chiral Perturbation Theory in finite volumes, when the volume is not sufficiently large compared with the Compton wavelength of the lighter pions. 

Clearly the methods developed here can be used for the computation of other correlation functions \cite{us_future}.

\section*{Acknowledgements}

We wish to thank  C.~Haefeli, M.~Laine, S. ~Necco and  C.~Pena for useful discussions on the topics of this work and a critical reading of the manuscript. We also thank S.~Aoki, S.~Hashimoto and T.~Degrand for informing us of their interest in this calculation. F.B. acknowledges the finantial support of the FPU grant AP2005-5201. This work was also partially supported by the Spanish CICYT (Project No.\ FPA2004-00996 and FPA2005-01678), by the Generalitat Valenciana (Project No.\ GRUPOS03-13) and the Integrated Grant HA2005-0066. 

\appendix
\renewcommand{\thesection}{\Alph{section}}
\renewcommand{\thesubsection}{\Alph{section}.\arabic{subsection}}
\renewcommand{\theequation}{\Alph{section}.\arabic{equation}}

%%%%%%%%%%%%%%%%%%%%%%%%%%%%%%%%%%%%%%%%%%%%%%%%%%%%%%%%%%%%%%%%%%%%%%%%

\newpage

\section{Calculation of the Jacobian}

The parametrization with some or all of the zero modes factorized, that we encounter in the mixed and $\epsilon$ regimes, has a non trivial Jacobian with respect to the $SU(N_f)$ Haar measure. Here we review first how the calculation of the Jacobian can be done for the $\epsilon$ regime parametrization at NLO. The method is easily extendable to the the mixed regime, so that we briefly mention which are the differences and give the results for the mixed regime too. In particular we show that the contribution  coming from the $\xi$ fields factorizes.\\
\\
A metric can be defined through:
\begin{equation}
\label{metric}
\langle dU^{\dagger}dU\rangle =g_{ab}dU_adU_b
\end{equation} 
so that the volume element $d\mu$ is obtained as:
\begin{equation}
d\mu=\sqrt{\det g} \bigwedge ^idU_i \,\,.
\end{equation}
The parenthesis $\langle . . . \rangle$ means here and in the following that both an integration and a trace are executed ($\int_V \frac{dx}{V}{\rm Tr}[...]$).\\
With the physical fields in the game, our metric matrix $g$ will be of the form
\begin{displaymath}
g=\begin{array}{|c|c|}
\hline
g_1 & g_2 \\
\hline
g_3 & g_4 \\
\hline
\end{array}
\end{displaymath}
and we will see that $g_1$ contains elements of order 1, $g_2=g_3^T$ contain elements of order $\epsilon$, and $g_4=1+O(\epsilon^2)$.\\
At next to leading order we are lead by the see saw formula to consider the matrix
\begin{displaymath}
g=\begin{array}{|c|c|}
\hline
g_1 & 0 \\
\hline
0 & g_4-g_2^Tg_1^{-1}g_2 \\
\hline
\end{array}
\end{displaymath}
that has the same eigenvalues and eigenvectors of $g$ to $\epsilon^2$ order. \\
We will use the parametrization:
\be
\label{param}
U=U_0U_{\xi}=e^{\frac{2i\phi}{F}}e^{\frac{2i\xi}{F}}
\ee
and the expansions:
$$
\phi=\phi^aT^a \qquad \xi (x)= \xi _{m_1}^{m_2}f_{m_1}(x)T^{m_2}\quad .
$$
Since $\xi$ only contain nonzero modes, the $f_{m_i}$ satisfy:
\begin{equation}
\label{nonzeromodes}
\int dx f_{m_i}(x)=0 \quad .
\end{equation}
Adding the constant function to the $f_{m_i}$ we obtain a complete set:
\begin{equation}
\sum_{m_i} f_{m_i}(x)f_{m_i}(y)=\delta(x-y)-1\quad .
\label{quasicomplete}
\end{equation}
The completeness relations for generators $T^a$ such that $Tr[T^aT^b]=\frac{1}{2}\delta^{ab}$ read:
\ba
\sum_{a=1}^{N^2-1}{\Tr}[T^aAT^aB]&=&-\frac{1}{2N}{\Tr}[AB]+\frac{1}{2}{\Tr}[A]{\rm Tr}[B]\\
\sum_{a=1}^{N^2-1}{\Tr}[T^aA]{\Tr}[T^aB]&=&-\frac{1}{2N}{\rm Tr}[A]{\Tr}[B]+\frac{1}{2}{\Tr}[AB]\\
\label{completeness}
\ea
and are valid if A, B are hermitian matrices. \\
Inserting (\ref{param}) in (\ref{metric}) we obtain:
\begin{equation}
\langle dU^{\dagger}dU\rangle =\langle U_{\xi}^{\dagger}dU_{\xi}U_0dU_0^{\dagger}+dU_{\xi}^{\dagger}U_{\xi}U_{\xi}^{\dagger}dU_{\xi}+U_0dU_0^{\dagger}dU_0U_0^{\dagger}+U_0^{\dagger}dU_{\xi}^{\dagger}U_{\xi}dU_0\rangle 
\label{UdUUdU}
\end{equation}
and note that blocks like $U^{\dagger}dU$ are easy to calculate since they are elements of the algebra.\\
The block containing  the $\xi$ fields gives:
%Now we use (\ref{defU}) at order $\epsilon ^3$ to calculate $U_{\xi}^{\dagger}dU_{\xi}$ in (\ref{UdUUdU}) at order $\epsilon ^2$, obtaining:
\begin{equation}
U_{\xi}^{\dagger}dU_{\xi}\simeq \frac{2id\xi}{F}+\frac{2}{F^2}(\xi d\xi-d\xi \xi)+\frac{4i}{F^3} (2\xi d\xi \xi-\xi^2d\xi-d\xi \xi^2)
\end{equation}
and consequently:
\begin{equation}
dU_{\xi}^{\dagger}U_{\xi}=(U_{\xi}^{\dagger}dU_{\xi})^{\dagger}=-U_{\xi}^{\dagger}dU_{\xi}\,\,.
\label{UdUxi}
\end{equation}
To calculate the block containing the zero modes we define the $M(y)$ matrix by:
\begin{equation}
M(y) \equiv e^{\frac{2iy}{F}\phi}e^{-\frac{2iy}{F}(\phi+d\phi)}
\end{equation}
and we see that $U_0dU_0^{\dagger}=M(1)-1$. $M$ is a solution for the Cauchy problem:
\begin{equation}
\frac{\partial M(y)}{\partial y} \simeq -\frac{2i}{F} d\phi ^a [e^{-\frac{2iy}{F}F^c\phi^c}]^{ab} T^b \qquad M(0)=1\,,
\label{difeq}
\end{equation}
if $F^a$ are the generators of the adjoint irrep ($[T^a,T^b]=if^{abc}T^c$, $[F^a]^{bc}=-if^{abc}$). Finally:
\begin{equation}
U_0dU_0^{\dagger}=-\left[ \frac{1-e^{-\frac{2i}{F}\tilde{\phi}}}{\tilde{\phi}}\right] ^{ab} d\phi^a T^b \equiv -A^{ab}d\phi ^aT^b=-dU_0U_0^{\dagger}
\label{UdU0}\end{equation} 
where $\tilde{\phi}=F^a\phi^a$. Last equality is a consequence of the reality of the structure constants.\\
Exploiting the orthonormality of the $f_{m_i}$, for $g_4$ one obtains:
\begin{equation}
\label{g4}
(g_4)_{m_1n_1}^{m_2n_2}=\langle dU_{\xi_{m_1}^{n_1}}^{\dagger}U_{\xi}U_{\xi}^{\dagger}dU_{\xi_{n_1}^{n_2}}\rangle = \frac{2}{F^2} (\delta_{m_1n_1}^{m_2n_2}+\frac{4}{3F^2}\int \frac{dx}{V}f_{n_1}f_{m_1}{\Tr}[T^{n_2}\xi T^{m_2}\xi-\xi^2T^{n_2} T^{m_2}])\,\,.
\end{equation}
At NLO $\det(1+a)\sim 1+\Tr[a]$ if the entries of $a$ are small. To take the trace of (\ref{g4}) one uses the completeness relations. The perturbative correction to the determinant is:
$$
\frac{2}{F^2}(\delta(0)-1)(-\frac{2N}{3F^2})\langle \xi^2 \rangle \quad .
$$
The addend proportional to $\delta(0)$ would be there even in the p regime but it is zero in dimensional regularization. And this is the explanation why we do not need to consider a measure term in the p regime.\\
The other correction at NLO comes from $Tr[g_2^Tg_1^{-1}g_2]$ and after straightorward calculations one sees that this amounts to 
\begin{equation}
Tr[g_2^Tg_1^{-1}g_2]=\frac{2}{F^2}\frac{2N}{F^2}  \langle \xi^2 \rangle \,\,.
\end{equation}
Combining these results we can calculate the measure:
\begin{equation}
d\mu \simeq  d\mu (\phi)_{Haar}d\xi J(\xi) \simeq d\mu (\phi)_{Haar}d\xi \frac{\sqrt{2}}{F} (1-\frac{2N}{3F^2}\langle \xi^2 \rangle)\,\,.\label{epsilonmeas}
\end{equation}
The same procedure can be applied to the mixed regime parametrizations. Of course the relations (\ref{nonzeromodes}) and (\ref{quasicomplete}) need to be modified properly. 
We obtained:
\begin{multline}
%d\mu \simeq d\mu (\phi)_{Haar}d\xi
 J(\xi)\simeq \frac{\sqrt{2}}{F}\left( 1-\frac{4}{3F^2V}\int dz  \sum _{m \in SU(N_v) } Tr[(T^m)^2\xi ^2-(T^m\xi)^2] \right.\nonumber \\
\left.-\frac{2}{F^2V^2}\int dz\, \int dw  \sum _{b,\,m \in SU(N_v) } Tr[T^b(\xi T^m-T^m\xi)](x)Tr[T^b(\xi T^m-T^m\xi)](y)\right)
%\label{measuremixed}
\end{multline}
in the factorization for full theory calculations,
\begin{multline}
%d\mu \simeq d\mu (\phi)_{Haar}d\xi 
J(\xi)\simeq \frac{\sqrt{2}}{F} \left(1-\frac{4}{3F^2V}\int dz  \sum _{m \in SU(N_v) \cup  T^{\eta} } Tr[(T^m)^2\xi ^2-(T^m\xi)^2] \right.\nonumber \\
\left.-\frac{2}{F^2V^2}\int dz\, \int dw  \sum _{b,\,m \in SU(N_v) \cup T^{\eta}} Tr[T^b(\xi T^m-T^m\xi)](x)Tr[T^b(\xi T^m-T^m\xi)](y)\right)
%\label{measuremixed2}
\end{multline}
in the factorization for PQ calculations.

%%%%%%%%%%%%%%%%%%%%%%%%%%% SECTION %%%%%%%%%%%%%%%%%%%%%%%%%%%%%%%%%%%%%%%
%
\section{Mixed-regime path integral}
\la{app:mixed}

In this appendix we discuss in detail the path integral in the mixed-regime. Starting with the factorization 
in eq.~(\ref{factor_mixed}), the path integral in the absence of sources can be written as
\ba
{\cal Z}_\nu = \int d\xi~J(\xi) ~\int_{SU(N_v)} d U_0 \int d \theta~e^{i \nu \theta} ~\exp\left( - {\cal S}\right) ,  
\ea
where ${\cal S}$ can be organized as an expansion in $\epsilon$ according to the power-counting  of eq.~(\ref{powerc}). The leading order is:
\ba
{\cal S}^{(0)} &=& - {\mu_s} \cos\left({\theta-\eta \over \Ns}\right) - {1 \over 2} \Tr\left[P_v \left(M  e^{i \eta/N_v} U_0 + U_0^\dagger e^{-i \eta/N_v} M\right)\right] \nonumber\\
&+&  \int d^4 x ~\left(\Tr\left[\partial_\mu \xi(x) \partial_\mu \xi(x)\right] + {M_{ss}^2}  \cos\left({\theta-\eta \over \Ns}\right) \Tr\left[ P_v \xi^2 \right]\right) ,
\ea
where the first term if ${\cal O}(\epsilon^{-2})$ and the others are all of ${\cal O}(1)$. 

Passing to the variables $\bar\theta\equiv \theta - \eta$ and ${\bar U}_0 \equiv e^{i \eta/N_v} U_0$:
\ba
{\cal Z}_\nu &=& \int d\xi~J(\xi) e^{- \int d^4 x \Tr\left[\partial_\mu \xi(x) \partial_\mu \xi(x)\right] }~\int_{U(N_v)} ~d {\bar U}_0 \det({\bar U}_0)^\nu    e^{  {1 \over 2} \Tr\left[P_v \left(M {\bar U}_0 + {\bar U}_0^\dagger M\right)\right]} e^{B(\xi,{\bar U}_0)} \nonumber\\
&&     \int  d \bar\theta~e^{i \nu \bar \theta}  ~e^{ \cos\left({\bar\theta\over \Ns}\right) \left( \mu_s + A(\xi)\right)} ,  
\ea
where 
\ba
A(\xi) \equiv M_{ss}^2 \int d^4 x \Tr\left[P_s \xi^2\right] + {\cal O} (\epsilon^2), 
\ea
and $B(\xi, \bar{U}_0)$ contains the ${\cal O}(\epsilon^2)$ terms of the original action that do no depend 
on $\bar\theta$. 

The integral over $\bar\theta$ is therefore of the form:
\ba
\int d \bar\theta  ~e^{F(\bar\theta)} ,  
\ea
where the real part of $F(\bar\theta)$ has a maximum at $\bar\theta=0$ and can be expanded 
in $\epsilon$ via a saddle point approximation. 
  Expanding $F(\bar\theta)$ around this maximum:
\ba
F(\bar\theta) = \left(\mu_s + A(\xi) \right)\left[ 1 - {1 \over 2} \left( {\bar\theta \over \Ns} \right)^2 + {\cal O}\left({\bar\theta\over \Ns}\right)^4\right] + i \nu \bar\theta, 
\ea
the integral can be rewritten as 
\ba
\int d\bar\theta ~e^{F(\bar\theta)} &=& e^{ \mu_s + A(\xi)} \int d\bar\theta e^{- {1 \over 2} \left( {\bar\theta \over \Ns} \right)^2 \left(\mu_s + A(\xi)\right)} ~e^{i \nu \bar\theta} \left(1 + {\cal O}(\bar\theta)^4  \right) \nonumber\\
& \simeq &e^{ \mu_s + A(\xi)}  {\sqrt{2\pi} \Ns \over \sqrt{\mu_s + A(\xi)}} ~e^{-{\nu^2 \Ns^2 \over 2 (\mu_s + A(\xi))}} \left(1 + {\cal O}\left({1 \over \mu_s+A(\xi)} \right) \right) \nonumber\\
&\simeq& C_\nu e^{A(\xi)} ~\left( 1 - {A(\xi) \over 2 \mu_s }+  {\cal O}(\epsilon^4 ) \right). 
\ea
Therefore at the order we need to go, the path integral can be written as:
\ba
{\cal Z}_\nu & \simeq& C_\nu \int d\xi~J(\xi) e^{- \int d^4 x \left(\Tr\left[\partial_\mu \xi(x) \partial_\mu \xi(x)\right] + M_{ss}^2  \Tr\left[ P_s \xi^2 \right]  \right)  } \nonumber\\
& & ~\int_{U(N_v)} ~d {\bar U}_0 \det({\bar U}_0)^\nu    e^{  {1 \over 2} \Tr\left[P_v \left(M {\bar U}_0 + {\bar U}_0^\dagger M\right)\right]}   \left ( 1 - {A(\xi)\over 2 \mu_s} +  B(\xi, {\bar U}_0) \right) .
\ea
 The LO path-integral factorizes into zero and non-zero mode contributions and the computation of correlation functions is then like in the $\epsilon$-regime.

\end{document}